\documentclass[12pt]{article}

\makeatletter
\def\singlespace{\def\baselinestretch{1}\@normalsize}

\usepackage{amssymb}

\usepackage{amsmath}
\usepackage{color}
\usepackage{epstopdf}

\usepackage{amsmath,amssymb}
\usepackage{booktabs}
\usepackage{graphicx}
\usepackage{mathrsfs}
\usepackage{verbatim}
\usepackage{times}
\usepackage{bm}
\usepackage{natbib}

\newtheorem{theorem}{Theorem}

\newtheorem{proposition}{Proposition}
\newtheorem{corollary}{Corollary}

\usepackage[plain,noend]{algorithm2e}

\makeatletter
\renewcommand{\algocf@captiontext}[2]{#1\algocf@typo. \AlCapFnt{}#2} 
\def\@algocf@capt@plain{top}
\renewcommand{\algocf@makecaption}[2]{%
  \addtolength{\hsize}{\algomargin}%
  \sbox\@tempboxa{\algocf@captiontext{#1}{#2}}%
  \ifdim\wd\@tempboxa >\hsize
    \hskip .5\algomargin%
    \parbox[t]{\hsize}{\algocf@captiontext{#1}{#2}}
  \else%
    \global\@minipagefalse%
    \hbox to\hsize{\box\@tempboxa}
  \fi%
  \addtolength{\hsize}{-\algomargin}%
}
\makeatother

\def\mb#1{\bm{#1}}
\newcommand{\bmm}{\mb{m}}

\newcommand{\bX}{\mb{X}}
\newcommand{\bY}{\mb{y}}

\newcommand{\bx}{\mb{x}}
\newcommand{\bv}{\mb{v}}

\newcommand{\bu}{\mb{U}}

\newcommand{\bbeta}{\mb{\beta}}

\newcommand{\btheta}{\mb{\theta}}
\newcommand{\bgamma}{\mb{\gamma}}
\newcommand{\bvarth}{\mb{\vartheta}}
\newcommand{\bdelta}{\mb{\Delta}}

\newcommand{\hbbeta}{\hat{\bbeta}}

\newcommand{\eps}{\varepsilon}
\newcommand{\beps}{\mb{\varepsilon}}
\newcommand{\bga}{\mb{\gamma}}

\def\wh{\widehat}
\def\Var{\mbox{var}}

\def\bu{\mb{u}}

\def\bv{\mb{v}}

\def\Cov{\mbox{cov}}

\def\pl{p_{\lambda}}

\def\beq{\begin{equation}}
\def\eeq{\end{equation}}
\def\beqn{\begin{eqnarray}}
\def\eeqn{\end{eqnarray}}
\def\beqnn{\begin{eqnarray*}}
\def\eeqnn{\end{eqnarray*}}

\def\sgn{\mbox{sgn}}
\def\mbm{\mb{m}}
\def\btheta{\mb{\theta}}
\def\bal{\mb{\alpha}}
\def\bE{\mb{E}}
\def\be{\mb{e}}
\def\bM{\mb{M}}
\def\bY{\mb{y}}

\graphicspath{{./art/}}

\numberwithin{equation}{section}
\renewcommand{\baselinestretch}{1.00}
\begin{document}

\title{Statistical Inference for Linear Mediation Models with High-dimensional Mediators and Application to
Studying Stock Reaction to COVID-19 Pandemic}

\author{Xu Guo$ ^a$, Runze Li$ ^b$, Jingyuan Liu$ ^c$ and Mudong Zeng$ ^b$\\
{\small $ ^a$School of Statistics, Beijing Normal University}\\
{\small Beijing, 100875, China}\\
{\small Email:xustat12@bnu.edu.cn}\\
{\small $ ^b$Department of Statistics, The Pennsylvania State University}\\
{\small University Park, PA 16802, USA.}\\
{\small Emails: rzli@psu.edu for R. Li;\quad muz149@psu.edu for M. Zeng}\\
{\small $ ^c$MOE Key Laboratory of Econometrics, Department of Statistics, School of Economics}\\
{\small Wang Yanan Institute for Studies in Economics and Fujian Key Lab of Statistics}\\
{\small Xiamen University, Xiamen, 361000, China}\\
{\small Email:jingyuan@xmu.edu.cn}
}

\date{}
\maketitle

\begin{abstract}
Mediation analysis draws increasing attention in
many scientific areas such as genomics, epidemiology and finance. In this paper,
we propose new statistical inference procedures for high dimensional mediation
models, in which both the outcome model and the mediator model are linear
with high dimensional mediators. Traditional procedures
for mediation analysis cannot be used to make statistical inference
for high dimensional linear mediation models due to high-dimensionality of the mediators. We propose an estimation procedure for
the indirect effects of the models via a partial penalized least squares
method, and further establish its theoretical properties. We further develop a
partial penalized Wald test on the indirect effects, and prove that the
proposed test has a $\chi^2$ limiting null distribution. We also propose an
$F$-type test for direct effects and show that the
proposed test asymptotically follows a $\chi^2$-distribution under null
hypothesis and a noncentral $\chi^2$-distribution under local alternatives.
Monte Carlo simulations are conducted to examine the finite sample performance
of the proposed tests and compare their performance with existing ones. We further apply
the newly proposed statistical inference procedures to study stock reaction to COVID-19 pandemic
via an empirical analysis of studying the mediation effects of financial metrics that bridge company's sector and stock return.

\end{abstract}



\renewcommand{\baselinestretch}{1.75}
\baselineskip=19pt

\section{Introduction}
Since the seminal work of \cite{Baron:Kenny:1986}, mediation
analysis has been used in various scientific research, such as
economics, psychology, pedagogy, and behavioral science
\citep{Conti:2016, Chernozhukov:2021,
Mackinnon:2008,Hayes:2013,Vanderweele:2015}. It is designed to
investigate the mechanisms whereby exposure variables affect an
outcome through intermediate variables, which are termed as
mediators. {For instance, in the field of policy evaluation, while
there certainly is no shortage of techniques assessing effects of
policies or other treatments on an outcome \citep{Imbens:2004,
Donald:2014, Athey:Imbens:Wager:2018, Ai:2021}, mediation analyses
move a step further to disentangle such effect into indirect
effects through mediators, such as certain economic indices, and
direct effects.} Numerous statistical inference procedures for
mediation models with low-dimensional mediators have been
extensively studied
\citep{Preacher:Hayes:2008,Vanderweele:Vansteelandt:2014}. See
\cite{Ten:Joffe:2010} and \cite{Preacher:2015} for brief reviews
about inferences under low-dimensional mediation models.

On account of modern data-collecting technology, mediation
analysis extends its territory to quantitative finance, genomics,
internet analysis, biomedical research, among other data-intensive
fields. This brings in high-dimensional mediators and requires
attention on high-dimensional mediation model (HDMM), where the
number of potential mediators is much larger than the sample size.
{Our work is motivated by such a high-dimensional mediation
structure when studying the effects of company's belonging sector
on stock return via influencing various financial metrics during
the COVID-19 period. Direct effects of sectors, as well as
financial statements, on stock performance have been extensively
studied in literature. See for instance
\cite{Fama:French:1993,Graham:2002,Callen:Segal:2004,Edirisinghe:Zhang:2008,Dimitropoulos:Asteriou:2009,Fama:French:2015,Khan:Khokhar:2015,Enke:Thawornwong:2005,Huang:2019}.
Yet as to be evidently shown by the empirical analysis in section
3.2, the companies' belonging sectors also significantly affect
stock returns indirectly through certain financial metrics in the
statements. In our analysis, 550 financial indexes are involved,
based on only 490 companies, resulting in high dimensional
mediators. }

{The high-dimensionality, on every account, poses both
computational and statistical challenges for carrying out
efficient mediation analysis. For instance, the traditional
structural equation modeling fails due to the rank-deficiency of
the observed covariance matrix. However, notwithstanding the high
dimensional mediation structure, the number of truly active
mediators is typically assumed small and less than the sample
size. This is referred to as the sparsity assumption in the
literature, although the sparsity pattern is unknown and thus to
be recovered. See, for example, \cite{Fan:Li:Zhang:Zou:2020} and
references therein.} Many existing methods in literature break
through such obstacle by utilizing the dimension reduction
techniques in regular linear models. For example,
\cite{Huang:Pan:2016} and \citet{Chen.etal:2018} adopted principal
components analysis to compress the dimensionality of mediators,
and applied bootstrap for inference. These methods are intuitive
and simple to implement, but lack theoretical justification about
asymptotic distributions of the test statistics. As an extension
of \cite{Huang:Pan:2016}, \cite{Zhao:Linqduist:Caffo:2020} further
introduced sparse principal component analysis to mediation
models. \cite{Zhang.etal:2016} used a two-stage technique with (a)
first screening out ``unimportant"
mediators, and then (b) applying {
existing procedures for the post-screened
outcome model.}
\cite{Zhou:Wang:Zhao:2020}  introduced debiased
penalized estimators for the direct and indirect effects, with theoretical
guarantees of the related tests. However, their method involves estimating high
dimensional matrices, leading to potentially unstable estimates and expensive
computation. Furthermore, imposing penalization on all parameters reduces the
efficiency of estimators, and hence tests. {
There are many developments on this
topic in the recent literature \citep{Chakrabortty:Nandy:Li:2018, Derkach:Pfeiffer:Chen:Sampson:2019,Song.etal:2020}.}

In this paper, we propose new statistical inference procedures for
HDMM. Statistical inference for high-dimensional data has
been an active research topic in the literature
\citep{Belloni2014, Zhang:Zhang:2014,van de
Geer:Buhlmann:Ritov:Dezeure:2014,Javanmard2014,
Shi:Song:Chen:Li:2019,FanY2020a,FanY2020b}. However, there are
much less work on statistical inference for HDMM. To our best
knowledge, \cite{Zhou:Wang:Zhao:2020} is the only one on testing
hypothesis on indirect effect with solid theoretical analysis. Our inference procedure on
indirect effect is distinguished from \cite{Zhou:Wang:Zhao:2020}
in that we observe the indirect effect in HDMM indeed is a low
dimensional parameter and is the difference between the total
effect and the direct effect in the HDMM. This motivates us to
estimate the total effect via least squares method and the direct
effect by partial penalized least squares method, and then
estimate the indirect effect by the difference between the
estimates of the total effect and the direct effect. We establish
the asymptotical normality of the indirect effect estimate and
further develop a Wald test for the indirect effect.

We estimate the direct effect in the HDMM by partial penalized
least squares method, and propose an $F$-type test for it. The
statistical inference on the direct effect essentially is the same
as statistical inference on low dimensional coefficients in
high-dimensional linear models. This topic has been studied under
the setting in which the covariate vector in the high-dimensional
linear models is fixed design \citep{Zhang:Zhang:2014,van de
Geer:Buhlmann:Ritov:Dezeure:2014, Shi:Song:Chen:Li:2019}.
Due to the nature of HDMM, the design matrix in HDMM must be random rather than fixed since mediators
are random. {Thus, the statistical setting studied in this paper is different from the one in
\cite{Shi:Song:Chen:Li:2019},
in which the covariate vector is assumed to be fixed design.}
We study the asymptotical property of the proposed estimator in the random-design setting.
The random design imposes challenges in
deriving the rate of convergence and asymptotical normality of the partial
penalized least squares estimates. Under mild regularity conditions, we prove
the sparsity and establish the rate of convergence of the partial penalized least squares
estimate. We further establish an asymptotical representation of the estimate.
Based on the asymptotical representation, we can easily derive the asymptotical
normality of the estimate and derive the asymptotical distributions of the
proposed test for the direct effect under null hypothesis and under
local alternative.

We show that the proposed estimate of indirect effect is asymptotically more
efficient than the one proposed in \cite{Zhou:Wang:Zhao:2020}, and indeed is
asymptotically efficient under normality assumption. {This is because the debias step of debiased
Lasso inflates the asymptotical variance of the resulting estimate}.
We conduct Monte
Carlo simulation studies to assess the finite sample performance of the proposed
estimate in terms of bias and variance and to examine Type I error and
power of the proposed test. We also conduct numerical comparisons among the
proposed estimate, the oracle estimate and the estimate proposed in \cite{Zhou:Wang:Zhao:2020}. Our numerical comparison indicates that the proposed estimate
performs as well as the oracle one, and outperforms the estimate proposed by
\cite{Zhou:Wang:Zhao:2020}.

{We  utilize the proposed method to study the mediator role of
financial metrics that bridge company's sector and stock return.
We select six financial metrics out of all the 550 that indeed
mediate the pathways linking company sector and stock return, with
interestingly and informatively financial interpretations. We also
compare the metrics selected using our data during the COVID-19
period and those classical findings in existing works, including
\cite{Fama:French:2015}, \cite{Edirisinghe:Zhang:2008}, among
others. We indeed discover some unique patterns and features due
to the pandemic. Moreover, according to the proposed tests for
effects of sector, both its direct effect and indirect effect via
financial metrics are statistically significant. Therefore,
evaluating the selected financial metrics, as well as the sector
information, might help investors to make wiser investment
decisions and choose stocks especially during the pandemic.

The rest of this paper is organized as follows. In section 2, we propose a new
statistical inference procedure for the indirect effect and establish its
theoretical properties. We also construct an $F$-type test for the direct effect. Section 3 presents numerical studies and a real data example. Conclusion and discussion are given in section 4. All proofs are presented in Appendix.}

\section{Tests of hypotheses on indirect and direct effects}
Consider the mediation models
\beqn
&&y= \bal_0^T\mbm + \bal_1^T \bx + \eps_1,\label{eqn2.1}\\
&&\mbm = \Gamma^T \bx+ \beps,\label{eqn2.2}
\eeqn
where $y$ is
the outcome, $\mbm$ is the $p$-dimensional mediator, $\bx$ is the
$q$-dimensional exposure variable, and $a^T$ denotes transpose of $a$. We
in this paper assume $p$ is high dimensional, while $q$ is fixed
and finite. Correspondingly, $\bal_0$ and $\bal_1$ are $p$- and
$q$-dimensional regression coefficient vectors, and $\Gamma$ is a
$q\times p$ coefficient matrix. Following the literature on high-dimensional mediation model
\citep{Zhang.etal:2016,van Kesteren:Oberski:2019,Zhou:Wang:Zhao:2020},
we impose a sparsity assumption that only a small proportion of entries in $\bal_0$ are nonzero.
This implies that the corresponding variables in $\mbm$ are actually relevant to $y$. Notably, from equation
(\ref{eqn2.2}), $\mbm$ must be random. We further assume that $\eps_1$ and
$\beps$ are independent random errors with
$\Var(\eps_1)=\sigma^2_1$ and $\Cov(\beps)=\Sigma^*$; 
$\eps_1$ is independent of $\mbm,\bx$, and $\beps$ is independent
of $\bx$.

Plugging (\ref{eqn2.2}) into (\ref{eqn2.1}) yields
\beq y= (\bbeta+\bal_1)^T\bx +
\eps_1+\eps_2=\bgamma^T\bx+\eps_3, \label{eqn2.6}
\eeq
where $\bbeta=\Gamma \alpha_0, \eps_2=\bal_0^T\beps$ with $\Var(\eps_2)=\sigma_2^2=\bal_0^T\Sigma^*
\bal_0, \bgamma=\bbeta+\bal_1$, and $\eps_3=\eps_1+\eps_2$ is
the total random error. {
Following the literature {\citep{Imai:Keele:Tingley:2010, Vanderweele:Vansteelandt:2014}}, we refer $\bbeta$ to the indirect effect of $\bx$ on $y$ mediated by $\mbm$, $\bal_1$ to the direct effect, and $\bgamma=\bal_1+\bbeta$ to the total effect. }
{A causal interpretation of $\bbeta$ and $\bal_1$ is briefly discussed in the Appendix.}

\subsection{
Estimating indirect and direct effects}
In practice, of interest is to test whether there exists significant {\color{black}(joint)} indirect effect or not.
This can be formulated as the following hypothesis testing problem
\beq
H_{0}: \bbeta = 0 \ \mbox{versus} \ H_{1}: \bbeta \ne 0. \label{eqn2.3}
\eeq
When both $p$ and $q$ are finite-dimensional, $\bbeta$ can be estimated
through $\hbbeta=\wh{\Gamma}\hat{\bal}_0$, where $\wh{\Gamma}$ and
$\hat{\bal}_0$ are $\sqrt{n}$-consistently estimated from models (\ref{eqn2.1})
and (\ref{eqn2.2}). That is, $\wh{\Gamma}=\Gamma + \bE_{\gamma}$ and
$\hat{\bal}_0=\bal_0 + \be_\alpha$, where $\bE_{\gamma}= O_P(1/\sqrt{n})$ and
$\be_{\alpha}=O_P(1/\sqrt{n})$ are estimation errors. Then
\beq \|\hat{\bbeta}
-\bbeta\|\leq \|\Gamma \be_{\alpha}\| + \|\bE_{\gamma} \bal_0\| +
\|\bE_{\gamma} \be_{\alpha}\|=O_P(1/\sqrt{n}), \label{eqn2.4}
\eeq
where
$\|\cdot\|$ stands for the Euclidean norm.

When $p$ is high-dimensional, however, the right-hand side of (\ref{eqn2.4}) is no longer
$O_P(1/\sqrt{n})$. This results in potentially non-ignorable estimation error
of $\hat{\bbeta}$. Moreover, $\bbeta$ is challenging to be estimated
through $\Gamma\bal_0$ as it involves estimation of a high-dimensional matrix
and a high-dimensional vector,
though, interestingly, $\bbeta=\Gamma\bal_0$ is $q$-dimensional, fixed and finite.

As a key observation from (\ref{eqn2.6}), the indirect effect
$\bbeta=\bgamma-\bal_1$, is the difference between the total effect and direct effect.
This motivates us to estimate $\bbeta$ by separately
estimating $\bgamma$ via (\ref{eqn2.6}) and $\bal_1$ via (\ref{eqn2.1}), respectively,
rather than estimating the high-dimensional $\Gamma$ and $\bal_0$.

Suppose that $\{\mbm_i, \bx_i, y_i\}$, $i=1,\cdots, n$ is a random sample from
(\ref{eqn2.1}) and (\ref{eqn2.2}). Let $\bY=(y_1,\cdots, y_n)^T$ and $\bX=(\bx_1,\cdots, \bx_n)^T$.
Then we estimate  $\bgamma$  by its least squares estimate
\beq
\hat\bgamma=(\bX^T\bX)^{-1}\bX^T\bY.\label{eqn2.7}
\eeq
While for the estimator of $\bal_1$, due to the high-dimensionality of $\bal_0$,
we propose the following partial penalized least squares method:
\beq
(\hat\bal_1,\hat\bal_0)=\arg\min_{\bal_1,\bal_0}\,\,\frac1{2n}\|\bY -\bM\bal_0- \bX \bal_1\|^2+\sum_{j=1}^p
\pl(|\alpha_{0j}|),
\label{eqn2.8}
\eeq
where $\bM=(\mbm_1,\cdots, \mbm_n)^T$ and $\pl(\cdot)$ is a
penalty function with a tuning parameter $\lambda$.
The regularization is only applied to the high-dimensional yet sparse $\bal_0$.
We opt not penalize $\bal_1$ to achieve local power
on the direct effect $\bal_1$ and the indirect effect $\bbeta$ under local alternatives.
See Theorem 2 and Corollary 1 below for more details.
Thus, our proposal is different from \cite{Zhou:Wang:Zhao:2020}, {in which
the central idea is to develop a debiased estimator not of $\bal_0$ or $\bbeta$,
but of $\tilde\Sigma_{XM}\bal_0$ with $\tilde\Sigma_{XM}=E[\bx\mbm^T]$.} 
This may lead to less efficient estimators due to debiasing, as discussed in the next subsection.


\subsection{Theoretical results}
In this section, we investigate statistical properties of the estimators. We first present some notations and assumptions. For the penalty function, it is assumed that $p_{\lambda}(t_0)$ is increasing and concave in $t_0\in [0,\infty)$, and has a continuous derivative $p'_{\lambda}(t_0)$ with
$p'_{\lambda}(0+)>0$. Denote $\rho(t_0,\lambda)=p_{\lambda}(t_0)/\lambda$ for $\lambda>0$.
Further, $\rho'(t_0,\lambda)$ is increasing in $\lambda\in(0,\infty)$ and
$\rho'(0+,\lambda)$ does not depend on $\lambda$.
Define
$\bar{\rho}(\mb{v},\lambda)=\{\sgn(v_1)\rho'(|v_1|,\lambda),\cdots,\sgn(v_l)\rho'(|v_l|,\lambda)\}^T$
for any vector $\mb{v}=(v_1,\cdots, v_l)^T$,
where $\sgn(\cdot)$ is the sign function. Define the local concavity
of $\rho(\cdot)$ at $\mb{v}$ as
$$\kappa(\rho,\mb{v},\lambda)=\lim_{\epsilon\rightarrow 0^+}\max_{1\leq j\leq l}
\sup_{t_1<t_2\in(|v_j|-\epsilon,|v_j|+\epsilon)}-\frac{\rho'(t_2,\lambda)-\rho'(t_1,\lambda)}{t_2-t_1}.$$

Let $\btheta=(\bal^T_1,\bal^T_0)^T$ and $\btheta_0=(\bal^{\star T}_{1},\bal^{\star T}_0)^T$,
the true value of $\btheta$. Further let
$\hat\btheta=(\hat\bal^T_1,\hat\bal^T_0)$ be the estimator of $\btheta_0$.
Denote $\mathcal{A}=\{j:\alpha^{\star}_{0j}\neq 0\}$, and $s=|\mathcal{A}|$ is the number of elements in $\mathcal{A}$. Moreover, $\bvarth=(\bal^T_1,\bal^T_{0,\mathcal{A}})^T$. And $\bvarth_0,\hat\bvarth$ are similarly defined. Let $\bM^j$ denote the $j$th column of $\bM$. Let $\bM_{\mathcal{A}}$ be the submatrix of $\bM$ formed by columns in $\mathcal{A}$. $\mbm_{i,\mathcal{A}}$ is the $i$th column of the matrix $\bM_{\mathcal{A}}^T$. Similarly, let $\bal^{\star}_{0,\mathcal{A}}$
be the subvector of $\bal^{\star}_0$ formed by elements in $\mathcal{A}$.
Define $\mathcal{A}^c=[1,\cdots,p]-\mathcal{A}$ as the complement set of $\mathcal{A}$.
Define $\mathcal{N}_0=\{\bm \delta\in R^s:\|\bm \delta-
\bal^{\star}_{0,\mathcal{A}}\|_2\leq d_n\}$.
Let $\Sigma_{MM}=E[\mbm_{\mathcal{A}}\mbm^T_{\mathcal{A}}]$,
$\Sigma_{MX}=E[\mbm_{\mathcal{A}}\bx^T]$, and $\Sigma_{XX}=E[\bx\bx^T]$.
Denote
\begin{eqnarray*}
\Sigma&=&\left(
  \begin{array}{ccc}
    \Sigma_{XX} & \Sigma_{XM}\\
    \Sigma_{MX}& \Sigma_{MM}\\
  \end{array}
\right).
\end{eqnarray*}
In this paper, for any vector $\mb{v}=(v_1,\cdots, v_l)^T$,
$\|\mb{v}\|_{\infty}=\max_i |v_i|$ and $\|\mb{v}\|_2=(\mb{v}^T\mb{v})^{1/2}$.
$\lambda_{\min}(A)$ and $\lambda_{\max}(A)$ denotes the minimum and maximum
eigenvalues of the matrix $A$, respectively. $\|A\|_{2,\infty}=\sup_{\mb{v}:\|\mb{v}\|_2=1}\|A\mb{v}\|_{\infty}$. Further $a\gg b$ means $\lim_{n\rightarrow
\infty} a/b=\infty$.
 We impose the following conditions:

\begin{itemize}
\item [A1.] $\lambda_{\min}(\Sigma)\geq c>0$, $\lambda_{\max}(\Sigma)=O(1)$, and $\|\bM^{T}_{\mathcal{A}^c}(\bX, \bM_{\mathcal{A}})\|_{2,\infty}=O_P(n)$.
\item [A2.] Let $d_n$ be the half minimum signal of $\bal^{\star}_{0,\mathcal{A}}$, i.e.
$d_n=\min_{j\in\mathcal{A}}|\alpha^{\star}_{0j}|/2$. Assume that $d_n\gg\lambda_{n}\gg\max\{\sqrt{s/n},\sqrt{\log p/n}\}$, $p'_{\lambda_n}(d_n)=o((ns)^{-1/2})$, $\lambda_n\kappa_0=o(1)$ where $\kappa_0=\max_{\delta\in\mathcal{N}_0}\kappa(\rho,\delta,\lambda_{n})$.
\item [A3.] For some $\varpi>2$, there exists a positive sequence $K_n$ such that $E[\|\mbm_{\mathcal{A}^c}\eps_{1}\|_{\infty}^{\varpi}]\leq K_n^{\varpi}$ and $K^2_{n}{\log p}/{n^{1-2/\varpi-\varsigma}}\rightarrow 0$ for some arbitrary small $\varsigma>0$. Further assume that $\max_{1\leq j\leq p+q}E(z^4_j)<C<\infty$, here $\mb{z}=(\mbm,\bx)$, $z_j$ is the $j$-th component of $\mb{z}$.
\end{itemize}

To emphasize the dependence on the sample size, in the above conditions and the Appendix,
we use $\lambda_n$ to denote the tuning parameter.
The first two conditions are mild and commonly assumed. See for instance \cite{Fan:Lv:2011}.
Condition A2 imposes a minimal signal condition on nonzero elements in $\bal_0$, but not on $\bal_1$.
Since our primary interest is to make statistical inference on direct effect $\bal_1$ and indirect effect
$\bbeta=\bga-\bal_1$, and $\bal_0$ may be treated as a nuisance parameter in this model. Thus, Condition A2
is reasonable in practice. Condition A3 is imposed for establishing sparsity result.
Compared with existing literature, A3 is very mild. In fact, to simplify the proof, some papers assume that all covariates are uniformly bounded - see for instance \cite{Wang:Wu:Li:2012}. Under bounded covariates condition, A3 reduces to $E(|\eps_{1}|^{\varpi})\leq C$ by taking $K_n$ as a constant. Furthermore, the dimension of $p$ is allowed to be an exponential order of the sample size $n$ according to conditions A2 and A3.

\begin{theorem}
Suppose that Conditions (A1)-(A3) hold, and $s=o(n^{1/2})$, then with probability tending to 1, $\hat\bal_0$ must satisfy (i) $\hat\bal_{0,\mathcal{A}^c}=0$.
(ii) $\|\hat\bal_{0,\mathcal{A}}-\bal^{\star}_{0,\mathcal{A}}\|_2=O_P(\sqrt{s/n})$. Let $\epsilon_1=(\eps_{11},\cdots,\eps_{n1})^T$. If further $s=o(n^{1/3})$, we obtain that
\begin{eqnarray*}
\sqrt n(\hat\bvarth-\bvarth_0)&=&\frac{1}{\sqrt n}\Sigma^{-1}\left(
  \begin{array}{ccc}
   \bX^T\epsilon_1\\
    \bM^T_{\mathcal{A}}\epsilon_1\\
  \end{array}
\right)+o_P(1).
\end{eqnarray*}
\end{theorem}
The above results provide the sparsity of $\hat\bal_0$,
the convergence rate of $\hat\bal_{0,\mathcal{A}}$ and the asymptotic representation
of $\hat\bvarth$, respectively. 

Based on the results in Theorem 1, we further obtain the following corollary:
\begin{corollary}
Suppose that Conditions (A1)-(A3) hold, and $s=o(n^{1/3})$, we have
\begin{eqnarray*}
\sqrt n(\hat\bal_1-\bal^{\star}_1)\rightarrow N(0, \sigma^2_1(\Sigma^{-1}_{XX}+B)),\ \mbox{and}\quad
\sqrt n(\hat\bbeta-\bbeta^{\star})\rightarrow N(0, \sigma^2_2\Sigma^{-1}_{XX}+\sigma^2_1B),
\end{eqnarray*}
where $B=\Sigma^{-1}_{XX}\Sigma_{XM}(\Sigma_{MM}-\Sigma_{MX}\Sigma_{XX}^{-1}\Sigma_{XM})^{-1}
\Sigma_{MX}\Sigma^{-1}_{XX}$,
and $\bbeta^{\star}$ is the true value of $\bbeta$.
\end{corollary}

This corollary presents the asymptotic normalities of
the estimators $\hat\bal_1$ and $\hat\bbeta$.
We next make theoretical comparison with the estimators in \cite{Zhou:Wang:Zhao:2020}.
Note that the asymptotic variance matrices of $\hat\bal^Z_1$ and $\hat\bbeta^Z$
in \cite{Zhou:Wang:Zhao:2020}  are
$\sigma^2_1(\Sigma^{-1}_{XX}+\tilde B)
$
and $\sigma^2_2\Sigma^{-1}_{XX}+\sigma^2_1\tilde B$, respectively, where
 $\tilde\Sigma_{MM}=E[\mbm\mbm^T], \tilde\Sigma_{MX}=E[\mbm\bx^T],$
 $\Sigma_{XX}=E[\bx\bx^T]$, and
$\tilde B=\Sigma^{-1}_{XX}\tilde \Sigma_{XM}(\tilde\Sigma_{MM}-\tilde\Sigma_{MX}\Sigma_{XX}^{-1}\tilde\Sigma_{XM})^{-1}\tilde\Sigma_{MX}\Sigma^{-1}_{XX}.
$
To show our proposed estimators are more efficient than those proposed in \cite{Zhou:Wang:Zhao:2020}, it suffices to show that $\tilde B>B$. Note that
$\Sigma^{-1}_{XX}+B=(I_q, 0_{q\times s})\Sigma^{-1}(I_q, 0_{q\times s})^T$, and
\begin{eqnarray*}
\Sigma^{-1}_{XX}+\tilde B&=&(I_q, 0_{q\times p})\left(
  \begin{array}{ccc}
    E[\bx\bx^T]& E[\bx\mbm^T]\\
    E[\mbm\bx^T]& E[\mbm\mbm^T]\\
  \end{array}\right)^{-1} (I_q, 0_{q\times p})^T\\
%
&=& (I_q, 0_{q\times s})
(\Sigma-E[\bx\mbm^T_{\mathcal{A}^c}]E[\mbm_{\mathcal{A}^c}\mbm^T_{\mathcal{A}^c}]^{-1}E[\mbm_{\mathcal{A}^c}\bx^T])^{-1}
(I_q, 0_{q\times s})^T.
%
\end{eqnarray*}
Thus, $\tilde B>B$ since
$(\Sigma-E[\bx\mbm^T_{\mathcal{A}^c}]E[\mbm_{\mathcal{A}^c}\mbm^T_{\mathcal{A}^c}]^{-1}E[\mbm_{\mathcal{A}^c}\bx^T])^{-1}>\Sigma^{-1}$.
Hence our proposed estimators are more efficient than those proposed in
\cite{Zhou:Wang:Zhao:2020}. {This should not be surprised because
the debias Lasso inflates its asymptotical variance in the debiased step for high-dimensional
linear model \citep{van de Geer:Buhlmann:Ritov:Dezeure:2014}. The proposed partial penalized least squares method
does not penalize $\bal_1$, and hence the debiased step becomes unnecessary.}

Under normality assumption that $\eps_{1}\sim N(0,\sigma^2_1)$
and $\beps\sim N(0,\Sigma^*)$,
it can be shown that our proposed estimators are indeed asymptotically efficient.
Under the normality assumption, the maximum likelihood estimator (MLE) of
$\bal_1, \bal_{0,\mathcal{A}}$ in the oracle model knowing $\bal_{0,\mathcal{A}^c}=0$
satisfies
\begin{eqnarray}\label{eqnB.1}
\left(\begin{array}{ccc}
    \bX^T(\bY-\bM_{\mathcal{A}}
\hat\bal^M_{0,\mathcal{A}}-\bX\hat\bal^M_1)\\
\bM^T_{\mathcal{A}}(\bY-\bM_{\mathcal{A}}\hat\bal^M_{0,\mathcal{A}}-\bX\hat\bal^M_1)\\
  \end{array}
\right)=0.
\end{eqnarray}
This implies that $\hat\bvarth^M=(\hat\bal^M_1,\hat\bal^M_{0,\mathcal{A}})$ has the same asymptotic distribution as $\hat\bvarth$.

Since the MLE of $\Gamma_{\mathcal{A}}$ is $\hat\Gamma^M_{\mathcal{A}}=(\bX^T\bX)^{-1}\bX^T\bM_{\mathcal{A}}$, the MLE of $\bbeta$ can be written as
\begin{eqnarray}\label{eqnB.2}
\hat\bbeta^M=\hat\Gamma^M_{\mathcal{A}}\hat\bal^M_{0,\mathcal{A}}=(\bX^T\bX)^{-1}\bX^T\bM_{\mathcal{A}}\hat\bal^M_{0,\mathcal{A}}.
\end{eqnarray}

By the definition of $\hat\bgamma$ and $\hat\bal_1$, it follows that 
\begin{eqnarray}\label{eqnB.3}
\hat\bbeta=\hat\bgamma-\hat\bal_1=(\bX^T\bX)^{-1}\bX^T\bY-(\bX^T\bX)^{-1}\bX^T(\bY-\bM\hat\bal_0)=
(\bX^T\bX)^{-1}\bX^T\bM\hat\bal_0.
\end{eqnarray}
Recall that Theorem 1 indicates that with probability tending to 1, $\hat\bal_{0,\mathcal{A}^c}=0$, and hence  \begin{eqnarray}\label{eqnB.4}
\hat\bbeta=(\bX^T\bX)^{-1}\bX^T\bM_{\mathcal{A}}\hat\bal_{0,\mathcal{A}}.
\end{eqnarray}
Note that $\hat\bal_{0,\mathcal{A}}$ and $\hat\bal^M_{0,\mathcal{A}}$ have the same asymptotic distribution.
Consequently, $\hat\bbeta$ has the
same asymptotic distribution as $\hat\bbeta^M$.
Thus it is asymptotically efficient.

\subsection{Test for indirect effect}

To form the test statistic for the indirect effect $\bbeta$, we first study its asymptotic variance matrix. Let $\hat{\mathcal{A}}=\{j:\hat\alpha_{0j}\neq 0\}.$ With probability tending to 1, we have $\hat{\mathcal{A}}=\mathcal{A}$.
Then the variance matrix $\Sigma$ and $\sigma^2_1$
can be estimated by the estimated sample version and the mean squared errors, respectively.
\begin{eqnarray*}
\hat{\Sigma}&=&\frac{1}{n}\left(
  \begin{array}{ccc}
    \bX^T\bX & \bX^T\bM_{\hat{\mathcal{A}}}\\
    \bM_{\hat{\mathcal{A}}}^T\bX& \bM_{\hat{\mathcal{A}}}^T\bM_{\hat{\mathcal{A}}} \\
  \end{array}
\right),
\quad \mbox{and}\quad
\hat\sigma^2_1=\frac{1}{n-\hat s-q}\|\bY -\bM\hat\bal_0-\bX\hat\bal_1\|^2,
\end{eqnarray*}
%
where $\hat s=|\hat{\mathcal{A}}|$. As is shown, $\hat\sigma^2_1=\sigma^2_1+o_P(1)$. In fact, when $s=o(n^{1/2})$, we have $\hat\sigma^2_1=\sigma^2_1+O_P(n^{-1/2})$.
Alternatively, we can estimate $\sigma^2_1$ using refitted cross-validation \citep{Fan:Guo:Hao:2012} or the scaled lasso \citep{Sun:Zhang:2013}.

As to $\sigma^2_2$, we first estimate
$\sigma^2=\Var(\eps_3)=\sigma^2_1+\sigma^2_2$ by
the classic least squares residual variance estimator $\hat\sigma^2$ based on
model (\ref{eqn2.6}). Thus $\hat\sigma^2_2=\hat\sigma^2-\hat\sigma^2_1$. In
practice, $\hat\sigma^2_1$ may sometimes be larger than $\hat\sigma^2$, where
we would simply set $\hat\sigma^2_2=0$. This is possible when no mediators are
relevant. That is, $\bal_0=0$, and hence $\sigma^2_2$ indeed equals zero.

According to Corollary 1,  the asymptotic variance matrices of $\hat\bal_1$ and $\hat\bbeta$ can be consistently estimated by:
\begin{eqnarray}\label{std_est}
\hat\sigma^2_1(I_q, 0_{q\times \hat s})\hat\Sigma^{-1}(I_q, 0_{q\times \hat s})^T;\,\, \hat\sigma^2_2\hat\Sigma^{-1}_{XX}+\hat\sigma^2_1[(I_q, 0_{q\times \hat s})\hat\Sigma^{-1}(I_q, 0_{q\times \hat s})^T-\hat\Sigma^{-1}_{XX}],
\end{eqnarray}
where $ \hat \Sigma_{XX}=\bX^T\bX/n$.
Then Wald test statistic for the hypotheses in (\ref{eqn2.3}) can be derived as
$$
S_n=n\hat\bbeta^T\left\{\hat\sigma^2_2\hat\Sigma^{-1}_{XX}+\hat\sigma^2_1[(I_q, 0_{q\times \hat s})\hat\Sigma^{-1}(I_q, 0_{q\times \hat s})^T-\hat\Sigma^{-1}_{XX}]\right\}^{-1}\hat\bbeta.
$$
Clearly, under $H_0$, $S_n\rightarrow \chi^2_q$, a chi-square random variable
with $q$ degrees of freedom.

To investigate the local power of $S_n$, we consider the local alternative hypotheses
$H_{1n}:\,\,\bbeta=\bdelta/\sqrt{n}$, where $\bdelta$ is a constant vector.
From Corollary 1, under such local alternative hypotheses,
$S_n\rightarrow \chi^2_{q}(\bdelta^T(\sigma^2_2\Sigma^{-1}_{XX}+\sigma^2_1B)^{-1}
\bdelta)$, a chi-square random variable with $q$ degrees of freedom
and noncentrality parameter $\bdelta^T(\sigma^2_2\Sigma^{-1}_{XX}+\sigma^2_1B)^{-1}
\bdelta$. Thus, $S_n$ can detect local effects that converge to 0 at root-$n$ rate.


\subsection{$F$-type Test on direct effect}

It is of interest to test the following hypothesis
\beq
H_{02}: \bal_1 = 0 \ \mbox{versus} \ H_{12}: \bal_1 \ne 0.
\label{eqn3.1}
\eeq
(\ref{eqn2.1}) and (\ref{eqn2.2}) are called complete or
full mediation models under $H_{02}$, while incomplete or partial mediation
models under $H_{12}$.

Testing the hypothesis in (\ref{eqn3.1}) essentially is to test low dimensional regression
coefficients in linear regression model (\ref{eqn2.1}).
This has been studied when the covariates in (\ref{eqn2.1}) are fixed design
\citep{Zhang:Zhang:2014,van de Geer:Buhlmann:Ritov:Dezeure:2014,Shi:Song:Chen:Li:2019}. Due to the nature of mediation model, the covariates in (\ref{eqn2.1})
are random design. The fixed-design assumption on $\mbm$ is inappropriate
in mediation models.

We will propose an $F$-type test for (\ref{eqn3.1}), and further show that the proposed
$F$-test asymptotically has a chi-square distribution with
$q$ degrees of freedom under $H_{02}$, and a noncentral chi-square distribution
with $q$ degrees of freedom under $H_{12}$. Similar to $F$-test, we need to
calculate the residual sum of squares (RSS) under the null and alternative hypotheses.
Under $H_{02}$, the penalized least squares function for model (\ref{eqn2.1}) becomes
\beq
\frac1{2n}\|\bY -\bM\bal_0\|^2+\sum_{j=1}^p \pl(|\alpha_{0j}|).
\label{eqn3.3}
\eeq
Denote by $\tilde{\bal}_0$ the resulting penalized least squares estimator. Then
the RSS under $H_{02}$ is $\mbox{RSS}_0=\|\bY -\bM\tilde\bal_0\|^2$. Under $H_{12}$,
we can estimate $\bal_0$ and $\bal_1$ by the partial penalized least squares method in
(\ref{eqn2.8}). Then we calculate
 $\mbox{RSS}_1=\|\bY -\bM\hat\bal_0-\bX\hat\bal_1\|^2$, the RSS under $H_{12}$.

The $F$-type test for hypothesis (\ref{eqn3.1}) is defined to be
\begin{equation}\label{eqn3.6}
    T_n= \frac{(\mbox{RSS}_0-\mbox{RSS}_1)}{\mbox{RSS}_1/(n-q)}.
\end{equation}
Theorem~\ref{thm2} below shows that the asymptotical null distribution of $T_n$
is a chi-square distribution with $q$ degrees of freedom.
To evaluate the local power of $T_n$ under local alternative hypotheses,
we impose the following assumption.
\begin{itemize}
\item[A4.] Consider local alternative hypotheses $H_{1n}: \bal_1=\mb{h}_n$.
Assume that $\|\mb{h}_n\|_2=O(\sqrt{1/n})$.
\end{itemize}
\begin{theorem}\label{thm2}
Suppose that Conditions (A1)-(A4) hold, and $s=o(n^{1/3})$. It follows that
\begin{eqnarray}\label{eqn3.7}
\sup_x|P(T_n\leq x)-P(\chi^2_{q}(n\mb{h}_n^T\Phi^{-1}\mb{h}_n/\sigma^2_1)\leq x)|\rightarrow 0.
\end{eqnarray}
Here $\Phi=(I_q, 0_{q\times s})\Sigma^{-1}(I_q, 0_{q\times s})^T$ and $\chi^2_{q}(n\mb{h}_n^T\Phi^{-1}\mb{h}_n/\sigma^2_1)$ is a chi square random variable with $q$ degrees of freedom
and noncentrality parameter $n\mb{h}_n^T\Phi^{-1}\mb{h}_n/\sigma^2_1$.
\end{theorem}

Theorem~\ref{thm2} implies that under  $H_{02}$,
$T_n$ asymptotically follows $\chi^2_q$ distribution, which does not depend
on any parameter in the model. This is similar to the Wilks phenomenon for likelihood
ratio test in classical statistical setting. In other words, the Wilks
phenomenon still holds in this high dimensional mediation model. Theorem~\ref{thm2} also
implies that $T_n$ can detect local alternatives that are distinct from the null
hypothesis at the rate of $1/\sqrt{n}$.

\subsection{Algorithm and tuning parameter selection}

To compute the partial penalized estimators $\hat\bal_1$ and $\hat\bbeta$, we apply the local linear approximation algorithm (LLA) in \cite{Zou:Li:2008} with the SCAD penalty in \cite{Fan:Li:2001},
$$p'_{\lambda}(t)=\lambda\{I(t\leq \lambda)
+\frac{(a\lambda-t)_{+}}{(a-1)\lambda}I(t>\lambda)\},$$
and set $a=3.7$. The tuning parameter $\lambda$ for our method is chosen based on the high-dimensional
BIC (HBIC) method in \cite{Wang:Kim:Li:2013}.
For a fixed regularization parameter $\lambda$, define
$$(\hat\bal_0^{\lambda}, \hat\bal_1^{\lambda})=\min_{\bal_0,\bal_1}\frac{1}{2n}\|\bY-\bM\bal_{0}-\bX\bal_1\|^2_2+\sum_{j=1}^p p_{\lambda}(|\alpha_{0,j}|).$$
The minimization of the  partial penalized least squares method can
be carried out as follows.
\begin{description}
\item 1. Get initial values for $\bal^{(0)}_0,\bal^{(0)}_1$ by minimizing a
partial $L_1$-penalized least squares:
$(\hat\bal_0^{(0)}, \hat\bal_1^{(0)})=\min_{\bal_0,\bal_1}
\frac{1}{2n}\|\bY-\bM\bal_{0}-\bX\bal_1\|^2_2+\lambda\sum_{j=1}^p |\alpha_{0,j}|.$
\item 2. Solve
$(\hat\bal_0^{(k+1)}, \hat\bal_1^{(k+1)})=\min_{\bal_0,\bal_1}
\frac{1}{2n}\|\bY-\bM\bal_{0}-\bX\bal_1\|^2_2+\sum_{j=1}^p p'_{\lambda}(|\alpha^{(k)}_{0,j}|)|\alpha_{0,j}|$
for $k=1,2,\cdots,$ until $\{(\hat\bal_0^{(k)}, \hat\bal_1^{(k)})\}$ converges.
\end{description}

In practice, we use a data-driven method to choose the
tuning parameter $\lambda$. Following \cite{Wang:Kim:Li:2013}, we use the HBIC criterion to choose $\lambda$. The HBIC score is defined as
%
$
\mbox{HBIC}(\lambda)=\log(\|\bY-\bM\bal_{0}-\bX\bal_1\|^2_2) + \mbox{df} \log(\log(n)) \log(p+q)/n,
$
where $\mbox{df}$ is the number of variables with nonzero coefficients in
$(\bal_0^T,\bal_1^T)^T$. Minimizing $\mbox{HIBIC}(\lambda)$ yields a selection of
$\lambda$.

%

\section{Numerical studies}
In this section, we examine the finite sample performance of the proposed procedures via
Monte Carlo simulation studies and illustrate the proposed procedure by a real data example.

\subsection{Simulation studies}\label{sec:sim1}
We first examine finite sample performances of the proposed
partial-penalization based test statistics, along with comparisons
with the oracle test statistics which know the true set
$\mathcal{A}=\{j:\alpha^{\star}_{0j}\neq 0\}$,
denoted as $S^O_n$ and $T^O_n$ as a benchmark, and the
debiased test statistics $S^Z_n$ and $T^Z_n$ in \cite{Zhou:Wang:Zhao:2020},
denoted by Zhou et al.'s method in the tables and figures in this section.
Note that \cite{Zhou:Wang:Zhao:2020}  focuses on the test of indirect effects. One can derive
a valid Wald test for direct effects based on the asymptotical normality
established in their paper.

\bigskip

\noindent{\bf Example 1}. In this example,  we set $n=300$, $q=1$, and $p=500$. $\bx\sim N(0,1)$ and $\mbm = \Gamma^T\bx +
\beps$, where $\beps\sim N(0, \Sigma^*)$ with $\Sigma^*$ being an AR
correlation structure. That is, the $(i,j)$-element of $\Sigma^*$  equals
$\rho^{|i-j|}$ and $\rho$ is set to be 0.5. Take
$\Gamma=c_1(\tau_1,\cdots,\tau_p)^T$, where $\tau_k=0.2k$ for $k=1,\cdots, 5$,
and when $k> 5$, $\tau_k$'s are independently generated from  $N(0,0.1^2)$. Set
$c_1=0$ to examine Type I error rate and $c_1=\pm0.1,\pm0.2,\cdots,\pm1$ for
power when testing the indirect effects.

We generate the response $y$ from model
$y=\bal_0^T\mbm + \bal_1^T\bx + \eps_1,$
where $\eps_1\sim N(0,0.5^2)$, $\bal_0=[1,0.8,0.6,0.4,0.2,0,\cdots 0]^T$ and $\bal_1=c_2$ is set in the same fashion as $c_1$.
The simulation results are based on 500 replications. The significance level is set to be $0.05$.

\begin{figure}
\centering
    \includegraphics[scale=0.65]{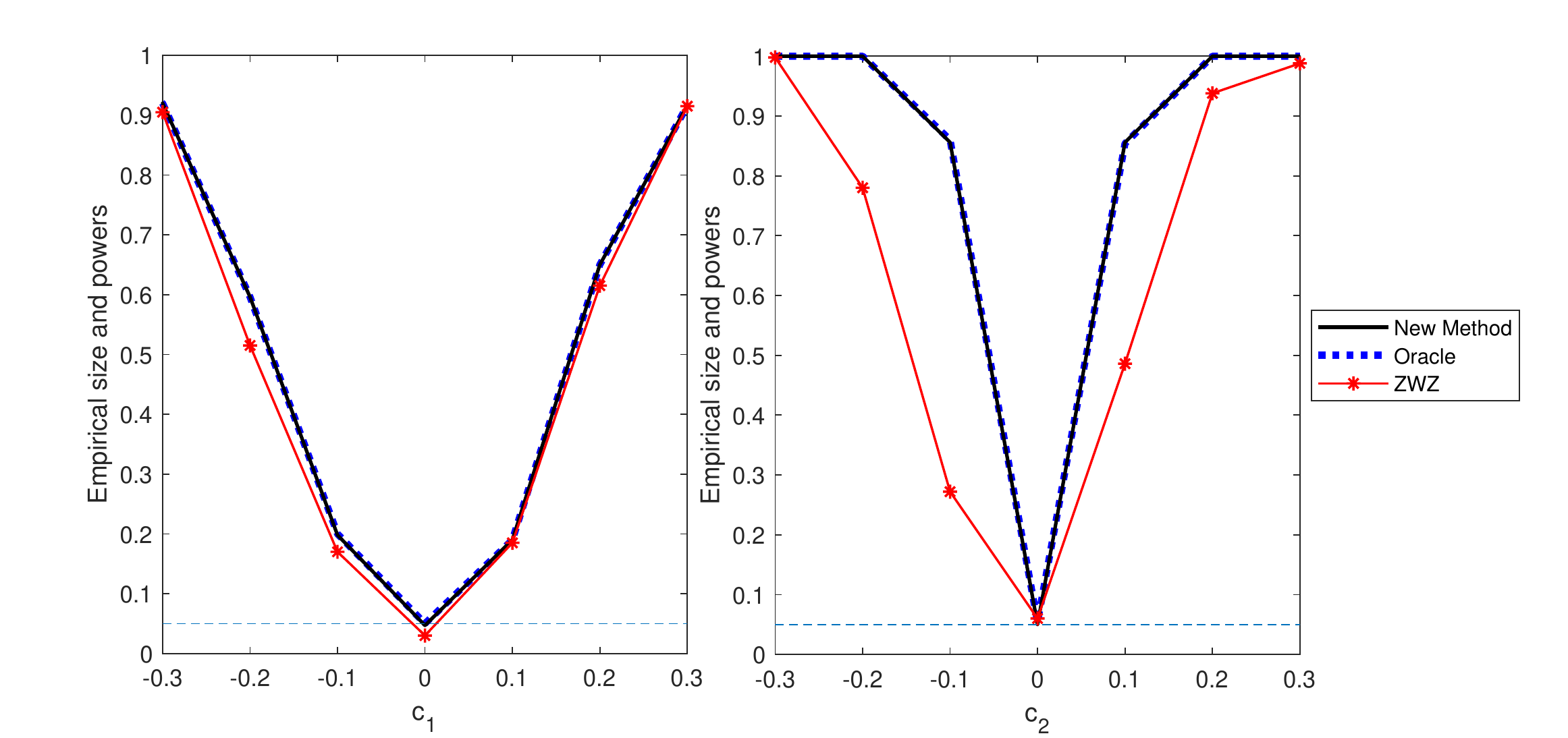}
\caption{Left panel is the empirical sizes and powers of $S_n, S^Z_n$
 and $S^O_n$ at level $\alpha = 0.05$ over 500 replications for testing indirect effect
 when $\alpha_1=0.5$.
 Solid line, dotted line and solid line marked by `*' represent the
 sizes and powers of $S_n, S^O_n$, and $S^Z_n$, respectively.
 Right panel is empirical sizes and powers of $T_n, T^Z_n$, and $T^O_n$
 at level $\alpha = 0.05$ over 500 replications for testing direct effect
 when $\beta=0.7$. The solid line, dotted line, and solid line marked by `*'
 represent the sizes and powers of $T_n, T^O_n$, and $T^Z_n$, respectively.}
\label{fig1}
\end{figure}

We first compare the performances of $S_n, S^O_n$ and $S^Z_n$ for testing the
indirect effect $\bbeta$.  We set $c_2=0.5$ and
$\bbeta=\Gamma\bal_0=1.4 c_1$. The left panel of Figure~\ref{fig1}
depicts power functions of the three tests
versus the values of $c_1$ over $[-0.3, 0.3]$. 
All the three tests gain larger powers as $|c_1|$ increases. $S_n$
performs as well as the oracle $S_n^O$, and is generally more powerful than
$S_n^Z$. For instance, when $c_1=-0.2$, the empirical power of $S^Z_n$ is
0.516, while the empirical powers of $S_n$ and  $S^O_n$ are 0.596. These
observations are in consistent with the theoretical results in Section 2.

Next, we turn to test the direct effect. Set $c_1=0.5$. And $c_2$ is taken from
$0,\pm0.1,\pm0.2, \cdots, \pm1$, where $c_2=0$ corresponds to the null hypothesis.
The right panel of Figure~\ref{fig1} depicts the power function of the three tests
versus the values of $c_2$ over $[-0.3, 0.3]$. 
The proposed test $T_n$ performs almost the
same as the oracle one, and is obviously more powerful than the test $T^Z_n$ proposed in \cite{Zhou:Wang:Zhao:2020}, whose power curve is asymmetric. 
In fact, when $c_2=-0.2$,
the empirical powers of our test statistic $T_n$ and the
oracle test $T^O_n$ are about 1, while that of $T^Z_n$  is only about
0.780.

\begin{singlespace}
\end{singlespace}

Furthermore, $T^Z_n$ performs unstably according to our simulation studies. To gain insight of this, we explore more on $\hat\bal^Z_1,\hat\bbeta^Z$.
The estimates $\hat\bal_1,\hat\bbeta$ and $\hat\bal^O_1,\hat\bbeta^O$ are reported in Table \ref{point_est_05} from which it can be
seen that the biases of $\hat\bal_1,\hat\bbeta$ and $\hat\bal^O_1,\hat\bbeta^O$
are very small, while $\hat\bal^Z_1$ has a large bias. This may be due to
that the direct effect $\bal_1$ is also penalized in \cite{Zhou:Wang:Zhao:2020} 's
estimation procedure based on scaled lasso. This makes sense only if the direct effect is
expected to be zero. As seen in Table \ref{point_est_05}, the bias of
$\hat\bal^Z_1$ is very small when $c_2=0$, yet inversely when $c_2\neq 0$.
Table \ref{point_est_05} also reports standard errors of corresponding estimates.
Both the proposed method and oracle outperform \cite{Zhou:Wang:Zhao:2020}, especially when estimating $\bal_1$.

\begin{singlespace}

\begin{table}[h]
\centering
\caption{Estimated biases and standard deviations (in parentheses) of
different methods with different $c_1$ and $c_2$. Except for $c_1$ and $c_2$,
the values in this table equals
 100 times of the actual ones.}
\scalebox{0.85}{
\begin{tabular}{cccccccc} \toprule
& & \multicolumn{2}{c}{New method}& \multicolumn{2}{c}{Oracle}&  \multicolumn{2}{c}{Zhou et al.'s method}\\
$c_1$   & $c_2$   & $\hat{\bal}_1$ & $\hat{\bbeta}$  & $\hat{\bal}^O_1$ &
$\hat{\bbeta}^O$  & $\hat{\bal}^Z_1$   & $\hat{\bbeta}^Z$     \\ \hline
-0.8 & 0.5 & $-0.23_{\tiny{(4.15)}}$ & $-0.22_{\tiny{(13.73)}}$ & $-0.11_{\tiny{(4.11)}}$ & $-0.35_{\tiny{(13.70)}}$ & $-11.77_{\tiny{(6.56)}}$ & $11.31_{\tiny{(14.05)}}$\\
-0.4 & 0.5 & $0.18_{\tiny{(3.13)}}$ & $-0.33_{\tiny{(11.98)}}$ & $0.25_{\tiny{(3.08)}}$ & $-0.40_{\tiny{(11.95)}}$ & $-3.49_{\tiny{(5.10)}}$ & $3.37_{\tiny{(12.20)}}$\\
0 & 0.5 & $-0.02_{\tiny{(2.99)}}$ & $0.39_{\tiny{(12.61)}}$ & $-0.00_{\tiny{(2.99)}}$ & $0.37_{\tiny{(12.63)}}$ & $-0.13_{\tiny{(8.65)}}$ & $0.47_{\tiny{(15.00)}}$\\
0.4 & 0.5 & $0.02_{\tiny{(3.15)}}$ & $0.08_{\tiny{(11.83)}}$ & $-0.02_{\tiny{(3.11)}}$ & $0.12_{\tiny{(11.81)}}$ & $-0.60_{\tiny{(5.31)}}$ & $0.77_{\tiny{(12.66)}}$\\
0.8 & 0.5 & $0.31_{\tiny{(3.79)}}$ & $0.26_{\tiny{(12.69)}}$ & $0.16_{\tiny{(3.72)}}$ & $0.42_{\tiny{(12.63)}}$ & $-1.57_{\tiny{(8.57)}}$ & $2.19_{\tiny{(15.05)}}$\\ \hline
0.5 & -0.8 & $0.16_{\tiny{(3.38)}}$ & $0.79_{\tiny{(11.62)}}$ & $0.11_{\tiny{(3.37)}}$ & $0.85_{\tiny{(11.64)}}$ & $16.37_{\tiny{(5.61)}}$ & $-7.63_{\tiny{(13.13)}}$\\
0.5 & -0.4 & $-0.01_{\tiny{(3.43)}}$ & $0.16_{\tiny{(12.58)}}$ & $-0.09_{\tiny{(3.36)}}$ & $0.26_{\tiny{(12.57)}}$ & $16.05_{\tiny{(4.00)}}$ & $-8.08_{\tiny{(13.64)}}$\\
0.5 & 0 & $0.10_{\tiny{(3.35)}}$ & $-0.15_{\tiny{(12.52)}}$ & $0.01_{\tiny{(3.33)}}$ & $-0.06_{\tiny{(12.52)}}$ & $0.66_{\tiny{(6.56)}}$ & $-0.71_{\tiny{(13.82)}}$\\
0.5 & 0.4 & $0.35_{\tiny{(3.39)}}$ & $0.01_{\tiny{(12.26)}}$ & $0.32_{\tiny{(3.37)}}$ & $0.04_{\tiny{(12.26)}}$ & $-0.96_{\tiny{(5.69)}}$ & $1.30_{\tiny{(13.10)}}$\\
0.5 & 0.8 & $0.13_{\tiny{(3.29)}}$ & $0.24_{\tiny{(12.10)}}$ & $0.05_{\tiny{(3.26)}}$ & $0.32_{\tiny{(12.17)}}$ & $-0.53_{\tiny{(5.58)}}$ & $0.84_{\tiny{(12.86)}}$\\
\bottomrule
\end{tabular}}
\label{point_est_05}
\end{table}
\end{singlespace}

To assess the accuracy of variance estimation of $\hat\bal_1$ and $\hat\bbeta$, Table~\ref{std_05} reports their estimated standard errors in two ways. As to each method - new, oracle and Zhou et al.'s method, the first column lists the empirical standard deviations of point estimates $\hat\bal_1$ or $\hat\bbeta$
over 500 replications (they are also recorded in parentheses of Table \ref{point_est_05}); for the second column, we estimate standard errors of $\hat\bal_1$ and $\hat\bbeta$ using formula (\ref{std_est}) in each simulation run, and reports the average together with standard deviations (in parentheses) over the 500 runs. Note that the R package ``freebird" \cite{Zhou:Wang:Zhao:2020} does not provide the estimated standard error of $\hat{\bal}_1$.
From Table~\ref{std_05}, for the new method and oracle, the standard errors estimated by Monte Carlo simulations are close to those calculated from formulas; while the two versions of \cite{Zhou:Wang:Zhao:2020}  depart more.

\begin{singlespace}
\begin{table}[h]
\centering
\caption{Estimated standard deviations and average estimated standard
errors with their standard deviations (in parentheses) over 500 replications
with different $c_1$ and $c_2$. Except for $c_1$ and $c_2$, the values in this
table equals
 100 times of the actual ones.}
\scalebox{0.85}{
\begin{tabular}{cc|cc|cc|cc|cc|cc}
\toprule
& & \multicolumn{4}{c}{Direct effect ($\hat{\alpha}_1$)}
&\multicolumn{6}{|c}{Indirect Effect ($\hat{\beta}$)} \\
\hline & & \multicolumn{2}{c|}{New method} & \multicolumn{2}{c|}{Oracle}
& \multicolumn{2}{c|}{New method} & \multicolumn{2}{c|}{Oracle} &  \multicolumn{2}{c}{Zhou et al.'s method}\\
\hline $c_1$   & $c_2$   & std & se(std)  & std & se(std)
      & std & se(std)   & std & se(std)     & std & se(std) \\
        \hline
-0.8 & 0.5 & 4.15 & $3.88_{\tiny{(0.23)}}$ & 4.11 & $3.89_{\tiny{(0.23)}}$ & 13.73 & $12.56_{\tiny{(0.72)}}$ & 13.70 & $12.56_{\tiny{(0.72)}}$ & 14.05 & $13.43_{\tiny{(1.03)}}$\\
-0.4 & 0.5 & 3.13 & $3.16_{\tiny{(0.18)}}$ & 3.08 & $3.17_{\tiny{(0.18)}}$ & 11.98 & $12.38_{\tiny{(0.73)}}$ & 11.95 & $12.38_{\tiny{(0.73)}}$ & 12.20 & $13.14_{\tiny{(0.85)}}$\\
0 & 0.5 & 2.99 & $2.90_{\tiny{(0.17)}}$ & 2.99 & $2.91_{\tiny{(0.17)}}$ & 12.61 & $12.26_{\tiny{(0.66)}}$ & 12.63 & $12.26_{\tiny{(0.66)}}$ & 15.00 & $13.12_{\tiny{(2.62)}}$\\
0.4 & 0.5 & 3.15 & $3.18_{\tiny{(0.18)}}$ & 3.11 & $3.19_{\tiny{(0.18)}}$ & 11.83 & $12.35_{\tiny{(0.71)}}$ & 11.81 & $12.35_{\tiny{(0.71)}}$ & 12.66 & $13.09_{\tiny{(0.82)}}$\\
0.8 & 0.5 & 3.79 & $3.88_{\tiny{(0.24)}}$ & 3.72 & $3.88_{\tiny{(0.23)}}$ & 12.69 & $12.47_{\tiny{(0.73)}}$ & 12.63 & $12.47_{\tiny{(0.73)}}$ & 15.05 & $13.37_{\tiny{(1.79)}}$\\ \hline
0.5 & -0.8 & 3.38 & $3.31_{\tiny{(0.19)}}$ & 3.37 & $3.32_{\tiny{(0.19)}}$ & 11.62 & $12.43_{\tiny{(0.71)}}$ & 11.64 & $12.42_{\tiny{(0.71)}}$ & 13.13 & $14.30_{\tiny{(0.76)}}$\\
0.5 & -0.4 & 3.43 & $3.30_{\tiny{(0.19)}}$ & 3.36 & $3.31_{\tiny{(0.20)}}$ & 12.58 & $12.30_{\tiny{(0.70)}}$ & 12.57 & $12.30_{\tiny{(0.70)}}$ & 13.64 & $13.19_{\tiny{(0.71)}}$\\
0.5 & 0 & 3.35 & $3.32_{\tiny{(0.18)}}$ & 3.33 & $3.33_{\tiny{(0.18)}}$ & 12.52 & $12.35_{\tiny{(0.75)}}$ & 12.52 & $12.34_{\tiny{(0.75)}}$ & 13.82 & $13.78_{\tiny{(3.73)}}$\\
0.5 & 0.4 & 3.39 & $3.32_{\tiny{(0.19)}}$ & 3.37 & $3.33_{\tiny{(0.19)}}$ & 12.26 & $12.39_{\tiny{(0.71)}}$ & 12.26 & $12.39_{\tiny{(0.71)}}$ & 13.10 & $13.14_{\tiny{(0.75)}}$\\
0.5 & 0.8 & 3.29 & $3.33_{\tiny{(0.20)}}$ & 3.26 & $3.34_{\tiny{(0.20)}}$ & 12.10 & $12.37_{\tiny{(0.74)}}$ & 12.17 & $12.37_{\tiny{(0.74)}}$ & 12.86 & $13.27_{\tiny{(1.31)}}$\\
\bottomrule
\end{tabular}}
\label{std_05}
\end{table}
\end{singlespace}

\begin{figure}[h]
\centering
\includegraphics[scale=0.5]{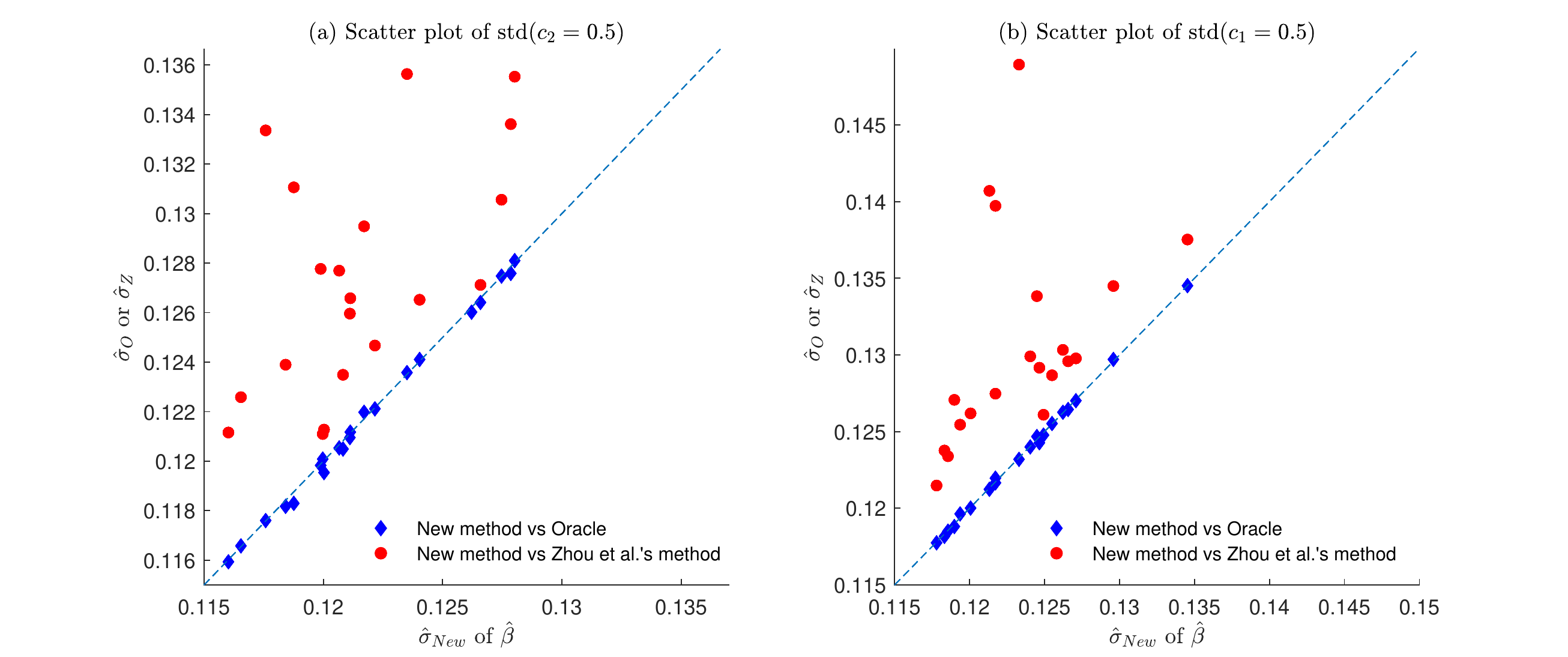}
\caption{Scatter plot of standard deviations of $\hat\bbeta$ over 500 point estimates
    by the new method ($x$-axis) and by oracle or Zhou et al.'s method ($y$-axis).
    Each dot (blue and red) corresponds each of the 21 different simulation
    settings - when holding $c_2 =0.5$, vary $c_1 = 0, \pm 0.1, \cdots ,\pm1$ in (a)
    and holding $c_1 =0.5$, vary $c_2 = 0, \pm 0.1, \cdots, \pm 1$ in (b).}
\label{fig:comp_std_05}
\end{figure}

Furthermore, Figure \ref{fig:comp_std_05} visually compares the standard deviations of $\hat\bbeta$ over 500 point estimates using the new method ($x$-axis) with those using oracle or Zhou et al.'s method ($y$-axis), respectively. Each blue diamond or red dot in the figure corresponds to each of the 21 different simulation settings - when holding $c_2 =0.5$, vary $c_1 = 0, \pm 0.1, \cdots ,\pm1$ in (a) and holding $c_1 =0.5$, vary $c_2 = 0, \pm 0.1, \cdots, \pm 1$ in (b). The figures imply that the estimated standard errors of the new method are close to oracle, and are generally smaller than those of Zhou et al.'s method. This in turn intuitively illustrates the precision of proposed estimators.

\begin{singlespace}
\end{singlespace}

Lastly, Table \ref{table:by:fig} reports the computing times, where the new method
is nearly 1000 times faster than Zhou et al.'s method. The proposed method is very fast and stable because initialized by LASSO estimator, LLA algorithm converges in one step.

\begin{singlespace}
\begin{table}
    \centering
    \caption{Comparison results of the average computing time
    (in seconds) over 500 replications.}
    \label{table:by:fig}
    \centering
    \begin{tabular}{cccc} \toprule
      $c_1$ & $c_2$ & New method & Zhou et al.'s method\\ \hline
      -0.8 & 0.5 & 1.38    & 1,207.88 \\
      -0.4 & 0.5 & 1.47   & 1,327.82 \\
      0 & 0.5 & 1.31 & 1,197.66 \\
      0.4 & 0.5 & 1.52  & 1,614.84 \\
      0.8 & 0.5 & 1.22    & 1,332.24 \\
      0.5 & -0.8 & 1.35    & 1,192.32\\
      0.5 & -0.4 & 1.33   & 1,329.48 \\
      0.5 & 0 & 1.48  & 1,544.23 \\
      0.5 & 0.4 & 1.50  & 1,790.34 \\
 \toprule
    \end{tabular}
\end{table}
\end{singlespace}

\bigskip

\begin{figure}[h]
\centering
\includegraphics[scale=0.6]{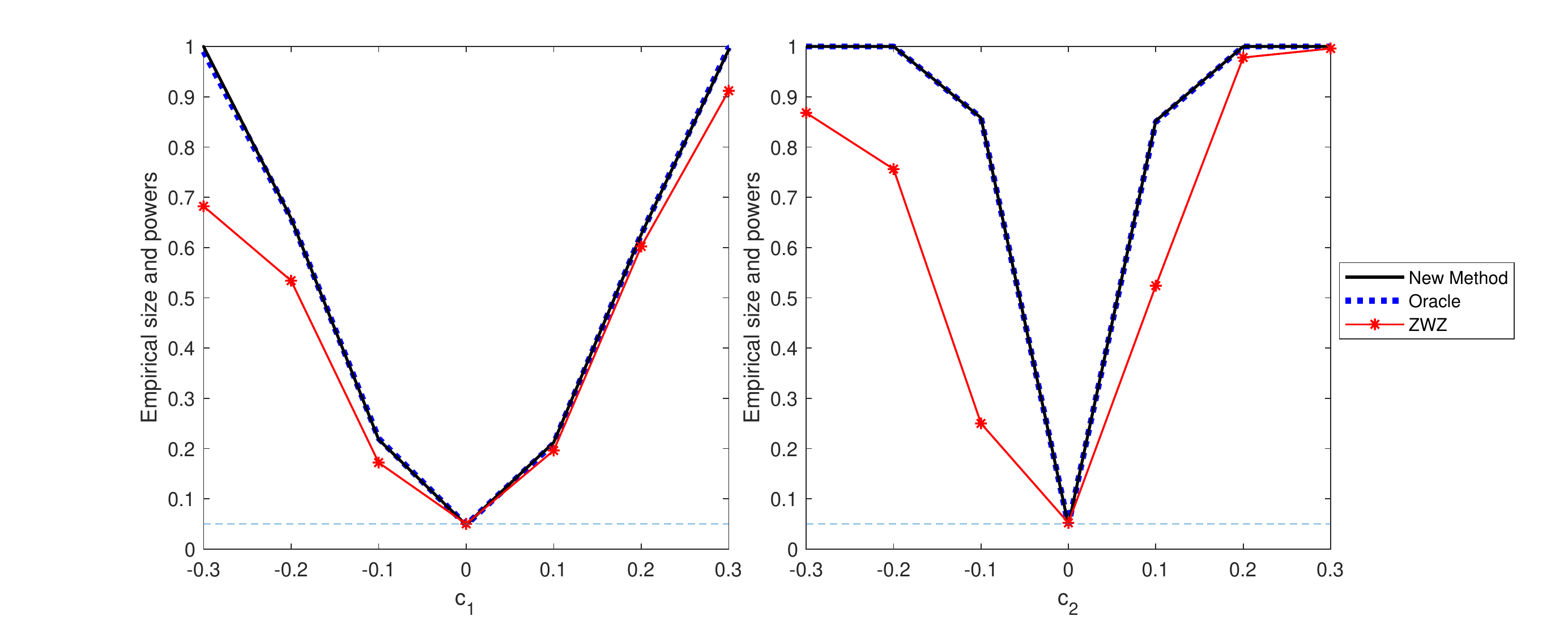}
    \caption{Left panel is empirical sizes and powers of $S_n, S^Z_n$
    and $S^O_n$ when $\varepsilon_1 \sim {t_6}/\sqrt{6}$ at
    level $\alpha = 0.05$ over 500 replications for testing indirect effect
    when  $\alpha_1 = 0.5$.
    Dotted line, solid line, and solid line marked by `*' represent the
    sizes and powers of $S_n, S^O_n$ and $S^Z_n$, respectively.
    Right panel is empirical sizes and powers of $T_n, T^Z_n$ and $T^O_n$
    for testing direct effect when  $\beta=0.7$.
    The dotted line, solid line, and solid line marked by `*' represent
    the sizes and powers of $T_n, T^O_n$ and $T^Z_n$, respectively.}
    \label{fig:Sn_t6}
\end{figure}

\noindent{\bf Example 2}. In this example, we examine the finite sample performances of proposed method when
heavy-tail errors are encountered. Specifically, assume now $\eps_1\sim
{t_6}/\sqrt{6}$. The multiplier $\sqrt{6}$ ensures the equality of variance of
$\eps_1$ to that when $\eps_1\sim N(0,0.5^2)$. All other settings are identical to those in Example 1.
We first investigate the performances of $S_n, S^O_n$ and
$S^Z_n$ for testing indirect effect $\bbeta$ via the left panel of
Figure~\ref{fig:Sn_t6}. The proposed test $S_n$ performs as well as the oracle
one  $S^O_n$ in terms of controlling Type-I error rate ($c_1=0$) and possessing
much larger power than $S^Z_n$ (when $c_1\neq 0$), especially when $c_1<0$.
Similar phenomenons are observed in the right penal of Figure~\ref{fig:Sn_t6}
when examining $T_n, T^O_n$ and $T^Z_n$. 
{The proposed test $T_n$ performs as well as
the oracle one, and is more powerful than the test $T^Z_n$.
In fact, when $c_2=-0.2$,
the empirical powers of our test statistic $T_n$ and the
oracle test $T^O_n$ are about 1, while that of $T^Z_n$  is only about
0.756.}
In addition, we also evaluate the accuracy and precision of $\hat\bal_1$
and $\hat\bbeta$ through Tables~\ref{tab:point_est_t6} and \ref{tab:std_t6}.
The overall pattern in these two tables with $\eps_1\sim {t_6}/\sqrt{6}$
is very similar to that
for $\eps_1\sim N(0,0.5^2)$. In sum, the proposed method retains its
validity for heavy-tailed error distributions.


\begin{singlespace}
\begin{table}[h]
\centering
\caption{Estimated biases and standard deviations
(in parentheses) of different methods with different
 $c_1$ and $c_2$ when $\varepsilon_1 \sim {t_6}/\sqrt{6}$.
Except for $c_1$ and $c_2$, the values in this table equals
100 times of the actual ones.}
\scalebox{1.0}{
\begin{tabular}{cccccccc} \bottomrule
& & \multicolumn{2}{c}{New method}& \multicolumn{2}{c}{Oracle}&  \multicolumn{2}{c}{Zhou et al.'s method}\\
$c_1$   & $c_2$   & $\hat{\bal}_1$ & $\hat{\bbeta}$  & $\hat{\bal}^O_1$ & $\hat{\bbeta}^O$  & $\hat{\bal}^Z_1$   & $\hat{\bbeta}^Z$     \\ \hline
-0.8 & 0.5 & $0.14_{\tiny{(4.06)}}$ & $-0.30_{\tiny{(12.46)}}$ & $0.22_{\tiny{(3.93)}}$ & $-0.38_{\tiny{(12.43)}}$ & $-13.93_{\tiny{(6.09)}}$ & $13.50_{\tiny{(12.84)}}$\\
-0.4 & 0.5 & $0.01_{\tiny{(1.93)}}$ & $-0.14_{\tiny{(6.24)}}$ & $0.06_{\tiny{(1.89)}}$ & $-0.19_{\tiny{(6.23)}}$ & $-3.34_{\tiny{(2.81)}}$ & $3.23_{\tiny{(6.43)}}$\\
0 & 0.5 & $0.16_{\tiny{(3.03)}}$ & $-0.36_{\tiny{(12.21)}}$ & $0.14_{\tiny{(3.01)}}$ & $-0.34_{\tiny{(12.21)}}$ & $-1.13_{\tiny{(4.68)}}$ & $0.86_{\tiny{(12.74)}}$\\
0.4 & 0.5 & $0.16_{\tiny{(3.29)}}$ & $-0.36_{\tiny{(12.30)}}$ & $0.09_{\tiny{(3.26)}}$ & $-0.28_{\tiny{(12.29)}}$ & $-0.77_{\tiny{(5.19)}}$ & $0.52_{\tiny{(13.01)}}$\\
0.8 & 0.5 & $0.28_{\tiny{(3.07)}}$ & $-0.26_{\tiny{(6.67)}}$ & $0.21_{\tiny{(3.02)}}$ & $-0.18_{\tiny{(6.63)}}$ & $0.75_{\tiny{(4.06)}}$ & $-0.70_{\tiny{(7.15)}}$\\ \hline
0.5 & -0.8 & $0.19_{\tiny{(3.44)}}$ & $-0.37_{\tiny{(12.34)}}$ & $0.10_{\tiny{(3.40)}}$ & $-0.28_{\tiny{(12.33)}}$ & $6.50_{\tiny{(5.61)}}$ & $-6.73_{\tiny{(12.89)}}$\\
0.5 & -0.4 & $0.16_{\tiny{(3.45)}}$ & $-0.32_{\tiny{(12.32)}}$ & $0.09_{\tiny{(3.41)}}$ & $-0.25_{\tiny{(12.30)}}$ & $5.92_{\tiny{(12.67)}}$ & $-6.16_{\tiny{(16.26)}}$\\
0.5 & 0 & $0.19_{\tiny{(3.42)}}$ & $-0.34_{\tiny{(12.34)}}$ & $0.09_{\tiny{(3.41)}}$ & $-0.25_{\tiny{(12.30)}}$ & $0.70_{\tiny{(4.56)}}$ & $-0.95_{\tiny{(12.95)}}$\\
0.5 & 0.4 & $0.20_{\tiny{(3.44)}}$ & $-0.39_{\tiny{(12.39)}}$ & $0.09_{\tiny{(3.41)}}$ & $-0.28_{\tiny{(12.33)}}$ & $-1.20_{\tiny{(5.30)}}$ & $0.93_{\tiny{(12.98)}}$\\
0.5 & 0.8 & $0.18_{\tiny{(3.44)}}$ & $-0.34_{\tiny{(12.32)}}$ & $0.09_{\tiny{(3.41)}}$ & $-0.25_{\tiny{(12.30)}}$ & $-1.17_{\tiny{(5.29)}}$ & $0.96_{\tiny{(13.07)}}$\\
\toprule
\end{tabular}}
\label{tab:point_est_t6}
\end{table}
\end{singlespace}

\begin{singlespace}
\begin{table}[h]
\centering
\caption{Estimated standard deviations and average estimated
standard errors with their standard deviations (in parentheses)
of different methods with different $c_1$ and $c_2$
when  $\varepsilon_1 \sim t_6/\sqrt{6}$.
Except for $c_1$ and $c_2$,  the values in this table equals
 100 times of the actual ones.}
\scalebox{0.8}{
\begin{tabular}{cc|cc|cc|cc|cc|cc}
\toprule
& & \multicolumn{4}{c}{Direct effect ($\hat{\alpha}_1$)}
&\multicolumn{6}{|c}{Indirect Effect ($\hat{\beta}$)} \\
\hline & & \multicolumn{2}{c|}{New method} & \multicolumn{2}{c|}{Oracle}
& \multicolumn{2}{c|}{New method} & \multicolumn{2}{c|}{Oracle} &  \multicolumn{2}{c}{Zhou et al.'s method}\\
\hline $c_1$   & $c_2$   & std & se(std)  & std & se(std)
      & std & se(std)   & std & se(std)     & std & se(std) \\
        \hline
-0.8 & 0.5 & 4.06 & $3.87_{\tiny{(0.28)}}$ & 3.93 & $3.88_{\tiny{(0.27)}}$ & 12.46 & $12.55_{\tiny{(0.70)}}$ & 12.43 & $12.55_{\tiny{(0.70)}}$ & 12.84 & $13.12_{\tiny{(0.90)}}$\\
-0.4 & 0.5 & 1.93 & $1.94_{\tiny{(0.14)}}$ & 1.89 & $1.95_{\tiny{(0.14)}}$ & 6.24 & $6.29_{\tiny{(0.33)}}$ & 6.23 & $6.29_{\tiny{(0.33)}}$ & 6.43 & $6.47_{\tiny{(0.36)}}$\\
0 & 0.5 & 3.03 & $2.91_{\tiny{(0.21)}}$ & 3.01 & $2.92_{\tiny{(0.21)}}$ & 12.21 & $12.29_{\tiny{(0.72)}}$ & 12.21 & $12.29_{\tiny{(0.72)}}$ & 12.74 & $12.80_{\tiny{(0.77)}}$\\
0.4 & 0.5 & 3.29 & $3.17_{\tiny{(0.23)}}$ & 3.26 & $3.18_{\tiny{(0.23)}}$ & 12.30 & $12.35_{\tiny{(0.72)}}$ & 12.29 & $12.35_{\tiny{(0.72)}}$ & 13.01 & $12.95_{\tiny{(0.82)}}$\\
0.8 & 0.5 & 3.07 & $2.92_{\tiny{(0.22)}}$ & 3.02 & $2.93_{\tiny{(0.22)}}$ & 6.67 & $6.66_{\tiny{(0.30)}}$ & 6.63 & $6.66_{\tiny{(0.30)}}$ & 7.15 & $6.57_{\tiny{(0.64)}}$\\ \hline
0.5 & -0.8 & 3.44 & $3.31_{\tiny{(0.24)}}$ & 3.40 & $3.32_{\tiny{(0.24)}}$ & 12.34 & $12.39_{\tiny{(0.71)}}$ & 12.33 & $12.39_{\tiny{(0.71)}}$ & 12.89 & $12.98_{\tiny{(0.74)}}$\\
0.5 & -0.4 & 3.45 & $3.31_{\tiny{(0.24)}}$ & 3.41 & $3.32_{\tiny{(0.24)}}$ & 12.32 & $12.39_{\tiny{(0.71)}}$ & 12.30 & $12.39_{\tiny{(0.71)}}$ & 16.26 & $13.01_{\tiny{(0.87)}}$\\
0.5 & 0 & 3.42 & $3.31_{\tiny{(0.24)}}$ & 3.41 & $3.32_{\tiny{(0.24)}}$ & 12.34 & $12.39_{\tiny{(0.71)}}$ & 12.30 & $12.39_{\tiny{(0.71)}}$ & 12.95 & $12.96_{\tiny{(0.74)}}$\\
0.5 & 0.4 & 3.44 & $3.31_{\tiny{(0.24)}}$ & 3.41 & $3.32_{\tiny{(0.24)}}$ & 12.39 & $12.39_{\tiny{(0.71)}}$ & 12.33 & $12.39_{\tiny{(0.71)}}$ & 12.98 & $12.98_{\tiny{(0.81)}}$\\
0.5 & 0.8 & 3.44 & $3.31_{\tiny{(0.24)}}$ & 3.41 & $3.32_{\tiny{(0.24)}}$ & 12.32 & $12.39_{\tiny{(0.71)}}$ & 12.30 & $12.39_{\tiny{(0.71)}}$ & 13.07 & $12.99_{\tiny{(0.86)}}$\\
\bottomrule
\end{tabular}
}
\label{tab:std_t6}
\end{table}
\end{singlespace}

\begin{singlespace}

\end{singlespace}

\subsection{Real data analysis}
We apply the proposed method to an empirical analysis to examine whether
financial statements items and metrics mediate the relationship between company sectors
and stock price recovery after COVID-19 pandemic outbreak. While investors and researchers have reached a consensus ages ago that stock returns highly rely on companies' belonging sectors, recent studies more focus on using financial statements or market conditions to predict stock returns. \cite{Fama:French:1993}'s pioneering proposal of the three-factor model started this era, which captures patterns of return using market return, firm size and book-to-market ratio factors. \cite{Callen:Segal:2004} showed that accruals, cash flow, growth in operating income significantly influence stocks return. \cite{Edirisinghe:Zhang:2007,Edirisinghe:Zhang:2008} developed a relative financial strength metric based on data envelopment analysis \citep{Farrell:1957,Charnes:1978}, and found that return on assets and solvency ratio has high correlation with stock price return. To enhance prediction accuracy, deep neural network and data mining techniques were developed, with model inputs as historical financial statements and  output as stock price return \citep{Enke:Thawornwong:2005,Huang:2019,Lee:2019}. Meanwhile, it is reasonable to hypothesize that companies' sectors affect stock performances via influencing the associated financial metrics. Few existing works, however, study the mediating effects of such financial metrics. Hence our analysis aims to fill in this gap, and use the proposed mediation analysis to select important financial metrics, as well as to test the direct and indirect effects of companies' sectors on returns.

In addition, we in this analysis are specifically interested in the stock performance of S\&P 500 component companies during the COVID-19 pandemic period. As is known, the outbreak of the COVID-19 dealt a shock to the U.S. economy with unprecedented speed, and the government had to take a lockdown to stop spread of virus. The lockdown took a toll in the U.S. economy: business were closed, millions of people lost jobs and the price of an oil futures contract fell below zero. The crisis spread to the U.S. stock market, dragging down the major index S\&P 500 by 33.92\%. To help businesses, households and the economy, the Federal Reserve and the White House launched various rescue programs and take measures to stabilize energy prices from the end of March, 2020. Therefore, all these events and measures led the U.S. stock market to a V-shape pattern, thanks to which, the general financial rules from classical literature may not directly apply any more.

Admittedly, a number of recent literature studied the economic reaction to COVID-19 pandemic from sector or company level data \citep{Ramelli:2020,Zhang:2020,Baker:2020,Gormsen:2020,Devito:2020}.
\cite{Thorbecke:2020} analyzed sector-specific and macroeconomic variables as contributing factors to stock return in COVID-19 downturn and found that idiosyncratic factors negatively affected energy and consumer cyclical sectors.
\cite{Hassan:2020} investigated companies' transcripts of quarterly earnings call from January to September 2020 to investigate senior management's and major market participants' opinions about future prospects. They discovered several important factors related to accounting and business fundamentals, including supply chain, production and operations and financing, that are highly associated with stock market recovery from COVID-19. However, these methods mainly rely on prior financial knowledge to select low dimensional data for modeling, while ignore important company level factors. Besides, these methods only consider the relation of stock return to either sector level or company level while failing to recognize that the company's financial plays a role in mediating stock sector effects to stock price return.
Therefore, we use the proposed method to study the financial statement items or metrics that mediate the relationship between firm sectors and stock performance in this special period. This work may then shed light on how to select valuable stocks during a pandemic or any adverse event likewise.

In the mediation models, the response is taken to be the stock return from its highest price before the pandemic in February, 2020 to April 30th, 2020. The closed price is adjusted for both dividends and splits.
The potential mediators in $\bmm$ are 550 accounting metrics from financial statements of associated companies, scratched from Yahoo Finance on April 30, 2020.
We obtain firms' annual reports from fiscal year 2015 to 2019 and the first three quarterly reports in 2019. 
We use the firms' latest annual report to compute financial metrics and use previous annual reports to compute average growth rate of each financial metrics. 
The exposure variables in $\bx$, are companies' sectors according to Global Industry Classification Standard (GICS) that are coded as dummy variables. GICS classifies companies into eleven sectors: basic materials, communication services, consumer cyclical, consumer defensive, energy, financial services, healthcare, industrials, real estate, technology and utilities. We set energy sector as baseline level.

Table \ref{tab:real_data1} presents the estimated direct and indirect effects of companies' sectors, together with their standard errors. We also calculate Wald's test for the indirect effect and generalized likelihood test for direct effect, with $p$-values smaller than $10^{-9}$ and $10^{-15}$, respectively, indicating both the direct and indirect effect are significant. As for direct effect, stocks in sectors such as healthcare and technology are more likely to outperform benchmark than ones from utilities sector. Furthermore, sectors influence the stocks performance partly through business operation reflected by selected financial metrics, and the indirect effects are significantly positive.

\begin{table}
\caption{\linespread{1.15}\small The estimated coefficients, standard errors,
test statistics values and $p$-values for real data.} \footnotesize
\begin{center}
\begin{tabular}{lcc|cc}
\hline \hline
Sectors & Direct effect & std & Indirect effect & std \\ \hline
Intercept               & -0.4634  &     0.1558  &  -0.5216  &       0.1016 \\
Basic materials         &  0.5725  &     0.2226  &   0.4450  &       0.1407 \\
Communication services  &  0.9231  &     0.2698  &   0.4227  &       0.1691 \\
Consumer cyclical       &  0.0793  &     0.1805  &   0.4154  &       0.1165 \\
Consumer defensive      &  0.9808  &     0.2087  &   0.6265  &       0.1386 \\
Financial services      &  0.1363  &     0.1844  &   0.3452  &       0.1206 \\
Healthcare              &  1.0176  &     0.1887  &   0.7601  &       0.1232 \\
Industrials             &  0.3658  &     0.1816  &   0.5899  &       0.1181 \\
Real.Estate             &  0.0736  &     0.2185  &   0.5010  &       0.1365 \\
Technology              &  0.6537  &     0.1823  &   0.7655  &       0.1203 \\
Utilities               &  0.6798  &     0.2121  &   0.3717  &       0.1343 \\\hline
$p$-value              &  $< 1\times 10^{-9}$ &  & $< 1 \times 10^{-15}$ \\ \hline
\end{tabular}\label{tab:real_data1}
\end{center}
\end{table}

The selected mediating metrics, their associated estimated coefficients in model (\ref{eqn2.1}), as well as their brief descriptions,  are presented in Table \ref{tab:real_data2}. These selected metrics are of their own significance. For instance,  the first three chosen metrics in Table \ref{tab:real_data2}, namely return on assets, gross margin and annual growth rate of operating income, reflect firms' revenue. Return on assets is an indicator of how well a firm utilizes its assets, by determining how profitable a firm is relative to its total assets. A firm with a higher return-on-assets value is preferred, as the firm squeezes more out of limited resources to make a profit. Gross margin is the portion of sales revenue a firm retains after subtracting costs of producing the goods it sells and the services it provides. It measures the gross profit of a firm. A firm that has higher gross margin is more likely to retain more profit for every dollar of good sold. Annual growth rate of operating income shows the firm's growth of generating operating income compared with previous year. Operating income measures the amount of profit realized from a business's operation, after deducting operating expenses such as wages, depreciation, and cost of goods sold. A firm with high growth of operating income can avoids unnecessary production costs, and improve core business efficiency. In a word, a firm with higher return on assets, gross margin and growing operating income is considered profitable,  and hence, is likely to attract investors.

On the other hand, both the average growth rate of quick ratio and debt to assets are indicators of financial leverage of a firm. Quick ratio  of a firm is defined as the dollar amount of liquid assets dividing that of current liabilities, where liquid assets are the portion of assets that can be quickly converted into cash with minimal impact on the price received in open market, while current liabilities are a firm's debts or obligations to be paid to creditors within one year. Thus a large quick ratio indicates that the firm is fully equipped with enough assets to be instantly liquidated to pay off its current liabilities. Debt to assets is the total amount of debt relative to assets owned by a firm. It reflects a firm's financial stability. Therefore, a firm with a higher quick ratio or a lower debt to assets might be more likely to survive when it is difficult to finance through borrowing and cover its debts, thus are more favorable to investors during the economy lockdown.

Lastly, receivables turnover quantifies a firm's effectiveness in collecting its receivables or money owed by clients. It shows how well a firm uses and manages the credit it extends to customers and how quickly that short-term debt is paid. Receivables turnover can be negative when net credit sale is negative because the client pre-pay for the product or service. A negative receivables turnover means that the firm are less susceptible to counter-party credit risk because it already receives the cash from its client before delivering the service or shipping out the product. This is especially important during liquidity dry periods when the clients may default or delay payment due to lack of cash. Therefore, a firm that has a negative receivables turnover is preferred.

On all accounts, one might incorporate the analysis results as reference when seeking for a stock portfolio during the financial crisis caused by pandemic. First, the sectors in `Healthcare', `Consumer defensive', `Communication service', `Utility' and `Technology' have the top five positive direct effects on stock return. In terms of the financial metrics, we may focus on those reported in Table~\ref{tab:real_data2} to filter stocks. For example, we shall select firms that have higher values in AGR operating income, gross margin, quick ratio, and return on assets but lower values in debt to assets and receivable turnover.

\begin{table}
\caption{\linespread{1.15}\small Selected importance mediators and their coefficients} \footnotesize
\begin{center}
\begin{tabular}{lcp{7.5cm}}
\hline \hline Selected mediator & Estimated coefficient (std) & Description\\ \hline
Return on assets  &  0.4246 (0.0379) & Net income divided by the total assets\\
Gross margin &  0.0841 (0.0393) & The difference between the revenue and cost of goods sold divided by revenue\\
AGR* Operating Income  &  0.1063 (0.0347)  & Revenues subtract the cost of goods sold and operating expenses\\
AGR* Quick ratio      &   0.1194 (0.0345) & Total current assets minus inventory divided by total current liabilities\\
Debt to assets   &-0.1209 (0.0369) & Total debts divided by total assets\\
Receivables turnover (days)  & -0.0947 (0.0346) & Average receivables divided by net credit sales times 360 days \\ \hline
\end{tabular}\label{tab:real_data2} \\
*  {AGR: average growth rate, calculated as the average of growth rates for the metrics from 2015 to 2019.}
\end{center}
\end{table}


Moreover, we compare our findings with those selected in established models. For instance, our method picks profitability factors like return on assets, which is also selected in \cite{Fama:French:2015}, as profitability is the core of a firm's stock performance. But we do not include metrics representing size of firm, valuation of stock price or investment that were covered by \cite{Fama:French:2015}. For firm size factor, there is no evidence that small-size firms recovered faster or slower than larger-size ones. For valuation of stock price factor, previous price valuation ratio changed significantly due to stock price change and is no longer reliable to predict future stock return. For investment factor, it is less important for a short-term stock price movement. Compared with \cite{Edirisinghe:Zhang:2008}, our method also picks profitability (return on assets), liquidity (quick ratio) and solvency (debt to assets) metrics, as in \cite{Edirisinghe:Zhang:2008}. During the crisis, a firm facing liquidity crunch could not access to credit. Therefore, a firm with sufficient cash and less debt is more easily to survive and less likely to be forced to liquidate valuable assets at unfavorable prices. And its stock would be safer and more attractive to investors. But we did not select metrics of earnings per share or about capital intensity as in \cite{Edirisinghe:Zhang:2008}. The lockdown dramatically changes a firm's revenue structure and capital allocation, and hence reduces predictive capability of these metrics to short-term recovery.

\section{Conclusion}
In this paper, we propose statistical inference procedures for the
indirect effects in high dimensional mediation model. We introduce
a partial penalized least squares method and study its statistical
properties under random design. We show that the proposed
estimators are more efficient than existing ones. We further
propose a partial penalized Wald test to detect the indirect
effect, with a $\chi^2$ limiting null distribution. In this paper,
we also propose an $F$-type test for the direct effect and reveal
Wilks phenomenon in the high-dimensional mediation model. {We
further utilize the proposed inference procedures to analyze the
mediation effects of various financial metrics on the relationship
between company's sector and the stock return.}

\section*{Acknowledgement}
Guo and Liu's research was supported by grants from National Natural Science Foundation of China
grants 12071038, 11701034, and 11771361, and Li and Zeng's research was supported by
National Science Foundation, DMS 1820702, 1953196 and 2015539.

\setcounter{equation}{0}
\setcounter{section}{0}
\setcounter{subsection}{0}
\renewcommand{\theequation}{A.\arabic{equation}}
\renewcommand{\thetable}{A.\arabic{table}}
\renewcommand{\thesubsection}{A.\arabic{subsection}}

\section*{Appendix}
\subsection{Proofs of Theorems}

Define
\begin{eqnarray*}
{Q}_n(\btheta)&=&\frac{1}{2n}\|\bY-\bM\bal_{0}-\bX\bal_1\|^2_2+\sum_{j=1}^p p_{\lambda}(|\alpha_{0,j}|).
\end{eqnarray*}

\vspace{3mm}

{\emph{Proof of Theorem 1}:} To enhance the readability, we divide the proof of Theorem 1 into three steps. In the first step,
we show that there exists a local
minimizer $\bar{\btheta}$ of $Q_n(\btheta)$ with the constraints $\bar{\bal}_{0,\mathcal{A}^c}=0$,
such that $\|\bar{\btheta}-\btheta_0\|_2=O_P(\sqrt{s/n})$. In the second step, we prove that
$\bar{\btheta}$ is indeed a local minimizer of $Q_n(\btheta)$. This implies $\hat\btheta=\bar{\btheta}$. In the final step, we derive the asymptotic expansion of $\hat\btheta$.

\vspace{3mm}

\emph{Step 1: Consistency in the $(s+q)$-dimensional subspace:} We first
constrain $Q_n(\btheta)$ on the $(s+q)$-dimensional subspace
of $\{\btheta\in R^{p+q}:\bal_{0,\mathcal{A}^c}=0\}$. This constrained partial penalized least squares function is given by
$$\bar{Q}_n(\bvarth)=\frac{1}{2n}\|\bY-\bM_{\mathcal{A}}\bm\delta-\bX\bal_1\|^2_2+\sum_{j=1}^sp_{\lambda}(|\delta_{j}|).$$
Here $\bvarth=(\bal^T_1,\bm \delta^T)^T$ and $\bm\delta=(\delta_1,\cdots,\delta_s)^T$. We now show that there exists a strict local
minimizer $\bar{\bvarth}$ of $\bar{Q}_n(\bvarth)$ such that $\|\bar\bvarth-\bvarth_0\|_2=O_P(\sqrt{s/n})$.
To this end, we consider an event
$$H_n=\{\min_{\bvarth\in\partial\mathcal{N}_{\tau}}\bar{Q}_n(\bvarth)>\bar{Q}_n(\bvarth_0)\}.$$
where $\mathcal{N}_{\tau}=\{\bvarth\in R^{s+q}:\|\bvarth-\bvarth_0\|_2\leq \tau\sqrt{s/n}\}$ with $\tau\in (0,\infty)$, and
$\partial\mathcal{N}_{\tau}$ denotes the boundary of the closed set $\mathcal{N}_{\tau}$. Clearly, on the event $H_n$, there exists a local minimizer of $\bar{Q}_n(\bvarth)$ in $\mathcal{N}_{\tau}$.
Thus, we only need to show that $P(H_n)\rightarrow 1$ as $n\rightarrow \infty$ when $\tau$ is large. To this aim, we next analyze the function
$\bar{Q}_n$ on the boundary $\partial\mathcal{N}_{\tau}$.

For any $\bvarth$, it follows from a second order Taylor's expansion that
\begin{eqnarray}\label{eqnA.1}
&&\bar{Q}_n(\bvarth)-\bar{Q}_n(\bvarth_0)=-(\bvarth-\bvarth_0)^T\mb{\nu}+\frac{1}{2}(\bvarth-\bvarth_0)^TD(\bvarth-\bvarth_0).
\end{eqnarray}

Here
\begin{eqnarray*}
\mb{\nu}&=&\left(
  \begin{array}{ccc}
  \frac{1}{n}\bX^T(\bY-\bM_{\mathcal{A}}
\bal^{\star}_{0,\mathcal{A}}-\bX\bal^{\star}_1)\\
\frac{1}{n}\bM^T_{\mathcal{A}}(\bY-\bM_{\mathcal{A}}
\bal^{\star}_{0,\mathcal{A}}-\bX\bal^{\star}_1)-\lambda_n\bar{\rho}(\bal^{\star}_{0,\mathcal{A}})\\
  \end{array}
\right)
,
\end{eqnarray*}
and
\begin{eqnarray*}
D&=&\frac{1}{n}\left(
  \begin{array}{ccc}
    \bX^T\bX & \bX^T\bM_{\mathcal{A}}\\
    \bM_{\mathcal{A}}^T\bX& \bM_{\mathcal{A}}^T\bM_{\mathcal{A}} \\
  \end{array}
\right)+\left(
  \begin{array}{ccc}
   0& 0 \\
    0& \Lambda(\bal^*_{0,\mathcal{A}})\\
  \end{array}
\right)\\
&=:&D_1+D_2.
\end{eqnarray*}
where $\bal^*_{0,\mathcal{A}}$ lies in the line segment jointing $\bm\delta$ and $\bal^{\star}_{0,\mathcal{A}}$, and $\Lambda(\bal^*_{0,\mathcal{A}})$ is a diagonal matrix with nonnegative diagonal elements. Clearly $\bal^*_{0,\mathcal{A}}\in\mathcal{N}_0$. By condition (A2), the maximum eigenvalue of $\Lambda(\bal^*_{0,\mathcal{A}})$ is upper bounded by $\lambda_n\kappa_0$.
Recall that
\begin{eqnarray*}
\Sigma&=&\left(
  \begin{array}{ccc}
    \Sigma_{XX} & \Sigma_{XM}\\
    \Sigma_{MX}& \Sigma_{MM}\\
  \end{array}
\right).
\end{eqnarray*}
Further note that
\begin{eqnarray*}
&&P(\|D_1-\Sigma\|_2\geq \eta)\leq \frac{1}{\eta^2}E[\|D_1-\Sigma\|^2_2]\leq \frac{cn}{\eta^2n^2}E[\sum_{i,j}^{s}[m_{1i}m_{1j}-E(m_{1i}m_{1j})]^2\\
&&+\sum_{i=1}^{s}\sum_{j=1}^q[m_{1i}x_{1j}-E(m_{1i}x_{1j})]^2
+\sum_{i,j}^{q}[x_{1i}x_{1j}-E(x_{1i}x_{1j})]^2]=\frac{cs^2}{\eta^2n}.
\end{eqnarray*}

Thus $\|D_1-\Sigma\|_2=O_P(s/\sqrt n)=o_P(1)$, when $s=o(n^{1/2})$.

Since $\lambda_{\min}(\Sigma)\geq c$ and $\lambda_n\kappa_0=o(1)$, we have:
\begin{eqnarray}\label{eqnA.2}
\lambda_{\min}(D)\geq \bar{c}>0.
\end{eqnarray}
 Consequently, we obtain
\begin{eqnarray*}
&&\min_{\bvarth\in\partial\mathcal{N}_{\tau}}\bar{Q}_n(\bvarth)-\bar{Q}_n(\bvarth_0)
\geq\min_{\bvarth\in\partial\mathcal{N}_{\tau}}\left(-\|\bvarth-\bvarth_0\|_2\|\mb{\nu}\|_2+\frac{1}{2}\|\bvarth-\bvarth_0\|^2_2\bar{c}\right)\\
&=&-\sqrt{\frac{s}{n}}\tau\|\mb{\nu}\|_2+\frac{1}{2}\frac{s}{n}\tau^2\bar{c}.
\end{eqnarray*}
By the Markov inequality, it entails that
\begin{eqnarray}\label{eqnA.3}
&&P(H_n)\geq P(\|\mb{\nu}\|_2\leq \frac{1}{2}\sqrt{\frac{s}{n}}\tau\bar{c})\geq 1-\frac{4nE\|\mb{\nu}\|^2_2}{s\tau^2\bar{c}^2}.
\end{eqnarray}
In the following, we aim to show that $E\|\mb{\nu}\|^2_2=O(s/n)$.

Note that
\begin{eqnarray*}
\mb{\nu}&=&\left(
  \begin{array}{ccc}
    \frac{1}{n}\bX^T\epsilon_1\\
\frac{1}{n}\bM^T_{\mathcal{A}}\epsilon_1\\
  \end{array}
\right)-\left(
  \begin{array}{ccc}
   0 \\
\lambda_n\bar{\rho}(\bal^{\star}_{0,\mathcal{A}})\\
  \end{array}
\right)=\mb{\nu}_1-\mb{\nu}_2,
\end{eqnarray*}
Then by condition (A1),
\begin{eqnarray*}
E\|\mb{\nu}_1\|^2_2&=&\frac{1}{n^2}\mbox{tr}\left[E\left(
  \begin{array}{ccc}
 \bX^T\epsilon_1\\
\bM^T_{\mathcal{A}}\epsilon_1\\
  \end{array}
\right)\left(
  \begin{array}{ccc}
\bX^T\epsilon_1\\
\bM^T_{\mathcal{A}}\epsilon_1\\
  \end{array}
\right)^T\right].\\
&=&\frac{\sigma^2_1}{n}\mbox{tr}(\Sigma)\leq \sigma^2_1\frac{s+q}{n}\lambda_{\max}(\Sigma)=O(\frac{s}{n}).
\end{eqnarray*}
It follows from the concavity of $\rho(\cdot)$, $d_n<|\alpha_{0j,\mathcal{A}}|$, and condition (A2) that:
\begin{eqnarray*}
\|\mb{\nu}_2\|^2_2\leq (s^{1/2}p'_{\lambda}(d_n))^2=o(\frac{1}{n}).
\end{eqnarray*}
Consequently, step 1 is completed.

\vspace{3mm}
\emph{Step 2: Sparsity:} According to Theorem 1 in Fan and Lv (2011), it suffices to show that with probability tending to 1, we have:
\begin{eqnarray}\label{eqnA.4}
\frac{1}{n}\|\bM^{T}_{\mathcal{A}^c}
(\bY-\bM\bar{\bal}_0-\bX\bar{\bal}_1)\|_{\infty}\ll\lambda_n.
\end{eqnarray}
Here $\bar{\btheta}=(\bar{\bal}_1^T, \bar{\bal}_0^T)^T$ satisfies that $\bar{\bal}_{0,\mathcal{A}^c}=0$
and $\|\bar{\btheta}-\btheta_0\|_2=O_P(\sqrt{s/n})$.
Note that
\begin{eqnarray}\label{eqnA.5}
\bM^{T}_{\mathcal{A}^c} (\bY-\bM\bar{\bal}_0-\bX\bar{\bal}_1)
=\bM^{T}_{\mathcal{A}^c}\epsilon_1-\bM^{T}_{\mathcal{A}^c}(\bX, \bM_{\mathcal{A}})(\bvarth-\bvarth_0).
\end{eqnarray}
For the second term,
\begin{eqnarray*}
\|\bM^{T}_{\mathcal{A}^c}(\bX, \bM_{\mathcal{A}})(\bvarth-\bvarth_0)\|_{\infty}\leq\|\bM^{T}_{\mathcal{A}^c}(\bX, \bM_{\mathcal{A}})\|_{2,\infty}\|\bvarth-\bvarth_0\|_2=O_P(\sqrt{ns}).
\end{eqnarray*}
%
%

Next we come to determine the rate of the first term $\|\bM^{T}_{\mathcal{A}^c}\epsilon_1\|_{\infty}$.

Let $a_n=n^{1/\varpi+\varsigma}K_{n}, b=\sqrt{Cn\log p}$ with $C$ being large enough and note that
\begin{align*}
m_{ij}\eps_{i1}&=m_{ij}\eps_{i1}I(|m_{ij}\eps_{i1}|\leq a_n)-E[m_{j}\eps_{1}{I}(|m_{j}\eps_{1}|\leq a_n)]\\
&+m_{ij}\eps_{i1}{I}(|m_{ij}\eps_{i1}|> a_n)-E[m_{j}\eps_{1}{I}(|m_{j}\eps_{1}|> a_n)]\\
&=:\epsilon_{ij,1}+\epsilon_{ij,2}.
\end{align*}

We have
\begin{align*}
&P\left(|\sum_{i=1}^n m_{ij}\eps_{i1}|>b,\,\,\mbox{for some}\,\,j\in\mathcal{A}^c\right)\\
&\leq P\left(|\sum_{i=1}^n \epsilon_{ij,1}|+|\sum_{i=1}^n \epsilon_{ij,2}|>b,\,\,\mbox{for some}\,\,j\in\mathcal{A}^c\right)\\
&\leq P\left(|\sum_{i=1}^n \epsilon_{ij,1}|>b/2,\,\,\mbox{for some}\,\,j\in\mathcal{A}^c\right)+P\left(|\sum_{i=1}^n \epsilon_{ij,2}|>b/2,\,\,\mbox{for some}\,\,j\in\mathcal{A}^c\right)\\
&=:P_1+P_2.
\end{align*}

Firstly consider the term $P_1$. Note that $\epsilon_{1j,1},\ldots,\epsilon_{nj,1}$ are independent centered random variables a.s. bounded by $2a_n$ in absolute value. Then the Bernstein inequality yields that
\begin{align*}
P_1&\leq 2(p-s)\max_j\exp\left\{-\frac{b^2/4}{2n E(\epsilon_{j,1}^2)+2\cdot 2a_n\cdot b/(2\cdot 3)}\right\}\\
&\leq 2p\max_j \exp\left\{-\frac{C\log p/4}{2E(\epsilon_{j,1}^2)+2a_n\sqrt{C\log p/n}/3}\right\}\rightarrow 0.
\end{align*}

Next we turn to consider $P_2$. First note that
\begin{align*}
P_2&\leq P\left(\sum\limits_{i=1}^{n}\max_j|m_{ij}\eps_{i1}|I(|m_{ij}\eps_{i1}|>a_n)+\max_j nE[|m_{j}\eps_{1}|{I}(|m_{j}\eps_{1}|>a_n)]>b/2\right)
\end{align*}

Further note that
\begin{align*}
E^2[|m_{j}\eps_{1}|{I}(|m_{j}\eps_{1}|>a_n)]\leq E[m^2_{j}\eps_{1}^2]P(|m_{j}\eps_{1}|>a_n)\leq
E[m^2_{j}\eps_{1}^2]\frac{E[|m_{j}\eps_{1}|^{\varpi}]}{a_n^{\varpi}}.
\end{align*}
We then conclude that
\begin{align*}
\max_jnE[|m_{j}\eps_{1}|{I}(|m_{j}\eps_{1}|>a_n)]\leq \max_jn\sqrt{\frac{E[m^2_{j}\eps_{1}^2]E[|m_{j}\eps_{1}|^{\varpi}]}{a_n^{\varpi}}}=o(\sqrt{n}).
\end{align*}
From this, we then have
\begin{align*}
P_2&\leq P\left(\sum\limits_{i=1}^{n}\max_j|m_{ij}\eps_{i1}|I(|m_{ij}\eps_{i1}|>a_n)>b/4\right)\\
&\leq P\left(\max_j|m_{ij}\eps_{i1}|> a_n\ \ \mbox{for some}\ i\right)\\
&\leq n\frac{E[\max_j|m_{j}\eps_{1}|^{\varpi}]}{a_n^{\varpi}}=o(1).
\end{align*}

Thus $\|\bM^{T}_{\mathcal{A}^c}\epsilon_1\|_{\infty}=O_P(\sqrt{n\log p})$.

Consequently, given condition A2, step 2 is finished.

\emph{Step 3: Asymptotic expansions:}
Steps 1 and 2 show that $\hat\bal_{0,\mathcal{A}^c}=0$ with probability tending to 1, and
further $\|\hat\bal_{0,\mathcal{A}}-\bal^{\star}_{0,\mathcal{A}}\|_2=O_P(\sqrt{s/n})$.

First denote
\begin{eqnarray}\label{eqnA.6}
\dot{L}(\bvarth_0)&=&\left(
  \begin{array}{ccc}
    \bX^T(\bY-\bM_{\mathcal{A}}
\bal^{\star}_{0,\mathcal{A}}-\bX\bal^{\star}_1)\\
\bM^T_{\mathcal{A}}(\bY-\bM_{\mathcal{A}}
\bal^{\star}_{0,\mathcal{A}}-\bX\bal^{\star}_1)\\
  \end{array}
\right)=\left(
  \begin{array}{ccc}
    \bX^T\epsilon_1\\
    \bM^T_{\mathcal{A}}\epsilon_1\\
  \end{array}
\right).
\end{eqnarray}
For $\hat\bvarth$, denote
\begin{eqnarray}\label{eqnA.7}
\dot{L}(\hat\bvarth)&=&\left(
  \begin{array}{ccc}
    \bX^T(\bY-\bM_{\mathcal{A}}
\hat\bal_{0,\mathcal{A}}-\bX\hat\bal_1)\\
\bM^T_{\mathcal{A}}(\bY-\bM_{\mathcal{A}}\hat\bal_{0,\mathcal{A}}-\bX\hat\bal_1)\\
  \end{array}
\right)=\left(
  \begin{array}{ccc}
    0\\
   n\lambda_n\bar{\rho}(\hat\bal_{0,\mathcal{A}})\\
  \end{array}
\right).
\end{eqnarray}

Notice that
$$\dot{L}(\bvarth_0)=\dot{L}(\hat\bvarth)+nD_1(\hat\bvarth-\bvarth_0).$$
Or equivalently we have
$$\frac{1}{\sqrt n}(\dot{L}(\bvarth_0)-\dot{L}(\hat\bvarth))=\Sigma\sqrt n(\hat\bvarth-\bvarth_0)+(D_1-\Sigma)\sqrt n(\hat\bvarth-\bvarth_0).$$
Recall that $\|D_1-\Sigma\|_2=O_P(s/\sqrt n)$, and
$\|\hat\bvarth-\bvarth_0\|=O_P(\sqrt{s/n})$. Then we have $$(D_1-\Sigma)\sqrt
n(\hat\bvarth-\bvarth_0)=o_P(1),$$ when $s=o(n^{1/3})$. Thus, we have
\begin{eqnarray*}
\sqrt n(\hat\bvarth-\bvarth_0)&=&\Sigma^{-1}\frac{1}{\sqrt n}(\dot{L}(\bvarth_0)-\dot{L}(\hat\bvarth))+o_P(1).
\end{eqnarray*}
Under condition (A2),
we have $\|\hat\bal_{0,\mathcal{A}}-\bal^{\star}_{0,\mathcal{A}}\|_{\infty}=O_P(\sqrt{s/n})\ll d_n$. This implies that
$$\min_{j\in\mathcal{A}}|\hat\alpha_{0j,
\mathcal{A}}|>\min_{j\in\mathcal{A}}|\alpha^{\star}_{0j,\mathcal{A}}|-d_n=d_n.$$
By the concavity of $p(\cdot)$ and condition (A2), we obtain that
$$\|n\lambda_n\bar{\rho}(\hat\bal_{0,\mathcal{A}})\|_2\leq ns^{1/2}p'_{\lambda_n}(d_n)=o(n^{1/2}).$$
Since $\lambda_{\max}(\Sigma^{-1})=O(1)$, it follows that
\begin{eqnarray}\label{eqnA.8}
\sqrt n(\hat\bvarth-\bvarth_0)&=&\Sigma^{-1}\frac{1}{\sqrt n}\dot{L}(\bvarth_0)+o_P(1).
\end{eqnarray}

{\emph{Proof of Corollary 1:}} Recall that
\begin{eqnarray*}
\Sigma^{-1}=\left(
  \begin{array}{ccc}
    \Sigma^{-1}_{XX}+\Sigma^{-1}_{XX}\Sigma_{XM}\Sigma_{MM.X}^{-1}\Sigma_{MX}\Sigma^{-1}_{XX} &  -\Sigma^{-1}_{XX}\Sigma_{XM}\Sigma_{MM.X}^{-1}\\
   -\Sigma_{MM.X}^{-1}\Sigma_{MX}\Sigma^{-1}_{XX}& \Sigma_{MM.X}^{-1}\\
  \end{array}
\right).
\end{eqnarray*}
Here $\Sigma_{MM.X}=\Sigma_{MM}-\Sigma_{MX}\Sigma^{-1}_{XX}\Sigma_{XM}$.

As a result, it follows that
\begin{eqnarray}\label{eqnA.9}
&&\sqrt n(\hat\bal_1-\bal^{\star}_1)=(I_{q\times q},0_{q\times s})\Sigma^{-1}\frac{1}{\sqrt n}\dot{L}(\bvarth_0)+o_P(1)\nonumber\\
&=&\frac{1}{\sqrt n}\Sigma^{-1}_{XX}\bX^T\epsilon_1
+\frac{1}{\sqrt n}\Sigma^{-1}_{XX}\Sigma_{XM}\Sigma_{MM.X}^{-1}(\Sigma_{MX}\Sigma^{-1}_{XX}\bX^T-\bM^T_{\mathcal{A}})\epsilon_1
+o_P(1).
\end{eqnarray}
The asymptotic variance matrix of $\hat\bal_1$ is
$$\sigma^2_1(I_{q\times q},0_{q\times s})\Sigma^{-1}(I_{q\times q},0_{q\times s})^T=\sigma^2_1\left(\Sigma^{-1}_{XX}+\Sigma^{-1}_{XX}\Sigma_{XM}\Sigma_{MM.X}^{-1}\Sigma_{MX}\Sigma^{-1}_{XX}\right).$$
Recall that
\begin{eqnarray}\label{eqnA.10}
\sqrt n(\hat\bgamma-\bgamma^{\star})&=&\frac{1}{\sqrt n}\Sigma^{-1}_{XX}\bX^T(\epsilon_1+\epsilon_2)+o_P(1).
\end{eqnarray}
Consequently we obtain that
\begin{eqnarray}\label{eqnA.11}
\sqrt n(\hat\bbeta-\bbeta^{\star})&=&\frac{1}{\sqrt n}\Sigma^{-1}_{XX}\bX^T\epsilon_2
+\frac{1}{\sqrt n}\Sigma^{-1}_{XX}\Sigma_{MX}\Sigma_{MM.X}^{-1}(\bM^T_{\mathcal{A}}-\Sigma_{MX}\Sigma^{-1}_{XX}\bX^T)\epsilon_1+o_P(1)\nonumber\\
&=&\frac{1}{\sqrt n}\Sigma^{-1}_{XX}\sum_{i=1}^n W_{1i}
+\frac{1}{\sqrt n}\Sigma^{-1}_{XX}\Sigma_{XM}\Sigma_{MM.X}^{-1}\sum_{i=1}^nW_{2i}+o_P(1).
\end{eqnarray}
Here $W_{1i}=\bx_i\eps_{2i}$ and
$W_{2i}=(\mbm_{i,\mathcal{A}}-\Sigma_{MX}\Sigma^{-1}_{XX}\bx_i){\eps_{1i}}$.

It is easy to show that
$E[W_{1i}]=E[\bx_i E(\eps_{2i}|\bx_i)]=0.$
Similarly, we have $E[W_{2i}]=E[(\mbm_{i,\mathcal{A}}-\Sigma_{MX}\Sigma^{-1}_{XX}\bx_i) E(\eps_{1i}|\bx_i,\mbm_{i,\mathcal{A}})]=0.$

Further we obtain that $\Var(W_{1i})=\sigma^2_2\Sigma_{XX}$,
$\Var(W_{2i})=\sigma^2_1\Sigma_{MM.X}$, and
\begin{eqnarray*}
\Cov(W_{1i},W_{2i})&=&E[\bx_i\eps_{2i}(\mbm_{i,\mathcal{A}}-\Sigma_{MX}\Sigma^{-1}_{XX}\bx_i){\eps_{1i}}]\\
&=&E[\bx_i\eps_{2i}(\mbm_{i,\mathcal{A}}-\Sigma_{MX}\Sigma^{-1}_{XX}\bx_i)E(\eps_{1i}|\bx_i,\mbm_{i,\mathcal{A}},\eps_{2i})]=0.
\end{eqnarray*}
As a result, it follows that
\begin{eqnarray}\label{eqnA.12}
\sqrt n(\hat\bbeta-\bbeta^{\star})\rightarrow N(0,\sigma^2_2\Sigma_{XX}^{-1}+\sigma^2_1\Sigma^{-1}_{XX}\Sigma_{XM}\Sigma_{MM.X}^{-1}\Sigma_{MX}\Sigma_{XX}^{-1}).
\end{eqnarray}

{\emph{Proof of Theorem 2:}} Similar to the arguments in the proof of
Theorem 1, we can also show that $\tilde\bal_{0,\mathcal{A}^c}=0$ with probability 1, and
further $\|\tilde\bal_{0,\mathcal{A}}-\bal^{\star}_{0,\mathcal{A}}\|_2=O_P(\sqrt{s/n})$.

Denote $\Delta\hat\bvarth=\hat\bvarth-\tilde\bvarth=(\Delta\hat\bvarth_1,\Delta\hat\bvarth_2)$ and \begin{eqnarray*}
\Sigma&=&\left(
  \begin{array}{ccc}
    \Sigma_{11} & \Sigma_{12}\\
    \Sigma_{21}& \Sigma_{22}\\
  \end{array}
\right),\,\,\,\Sigma^{-1}=\left(
  \begin{array}{ccc}
    \Sigma^{11} & \Sigma^{12}\\
    \Sigma^{21}& \Sigma^{22}\\
  \end{array}
\right).
\end{eqnarray*}

It is noted that
\begin{eqnarray}\label{eqnA.13}
\left(
\begin{array}{ccc}
0\\
n\lambda_n\bar{\rho}(\hat\bal_{0,\mathcal{A}})\\
\end{array}
\right)=\dot{L}(\hat\bvarth)=\dot{L}(\tilde\bvarth)-nD_1\Delta\hat\bvarth=\left(
  \begin{array}{ccc}
    L_1(\tilde\bvarth)\\
   n\lambda_n\bar{\rho}(\tilde\bal_{0,\mathcal{A}})\\
  \end{array}
\right)-\Sigma n\Delta\hat\bvarth-(D_1-\Sigma)n\Delta\hat\bvarth.
\end{eqnarray}
Here $D_1=\left(
  \begin{array}{ccc}
    \bX^T\bX & \bX^T\bM_{\mathcal{A}}\\
    \bM_{\mathcal{A}}^T\bX& \bM_{\mathcal{A}}^T\bM_{\mathcal{A}} \\
  \end{array}
\right)/n$.

From the proof of Theorem 1, it is known that
$\|n\lambda_n\bar{\rho}(\hat\bal_{0,\mathcal{A}})\|_2=o_P(n^{1/2})$ and
similarly
$\|n\lambda_n\bar{\rho}(\tilde\bal_{0,\mathcal{A}})\|_2=o_P(n^{1/2})$. Further
recall that $\|D_1-\Sigma\|_2=O_P(s/\sqrt n)$ and
$\|\Delta\hat\bvarth\|_2=\|(\hat\bvarth-\bvarth_0)-(\tilde\bvarth-\bvarth_0)\|_2=O_P(\sqrt{s/n})$.
Thus under condition that $s=o(n^{1/3})$, we have
\begin{eqnarray}\label{eqnA.14}
o_P(1)=\left(
  \begin{array}{ccc}
    \frac{1}{\sqrt n}L_1(\tilde\bvarth)\\
   0\\
  \end{array}
\right)-\left(
  \begin{array}{ccc}
    \Sigma_{11} & \Sigma_{12}\\
    \Sigma_{21}& \Sigma_{22}\\
  \end{array}
\right)\left(
  \begin{array}{ccc}
    \sqrt n\Delta\hat\bvarth_1\\
    \sqrt n\Delta\hat\bvarth_2\\
  \end{array}
\right),
\end{eqnarray}
from which we have
\begin{eqnarray}\label{eqnA.15}
\sqrt n\Delta\hat\bvarth_2=-\Sigma_{22}^{-1}\Sigma_{21}\sqrt n\Delta\hat\bvarth_1+o_P(1),
\ \mbox{and}\quad
\sqrt n\Delta\hat\bvarth_1=\Sigma^{11}\frac{1}{\sqrt n}L_1(\tilde\bvarth)+o_P(1).
\end{eqnarray}
Note that $\sqrt n\Delta\hat\bvarth_1=\sqrt n(\hat\bal_1-\bal^{\star}_1)+\sqrt n\mb{h}_n=O_P(1)$ from Corollary 1. Thus we get
$\sqrt n\Delta\hat\bvarth_2=O_P(1)$, which further implies that $\Delta\hat\bvarth_2^Tn\lambda_n\bar{\rho}(\tilde\bal_{0,\mathcal{A}})=o_P(1)$.

Now we are ready to investigate the asymptotic distribution of $T_n$. Under the
event $\hat\bal_{0,\mathcal{A}^c}=\tilde\bal_{0,\mathcal{A}^c}=0$ and recalling
equation (\ref{eqnA.15}), we can show that
\begin{eqnarray}\label{eqnA.16}
\mbox{RSS}_1-\mbox{RSS}_0&=&-2\Delta\hat\bvarth^T \dot{L}(\tilde\bvarth)+\Delta\hat\bvarth^T nD_1\Delta\hat\bvarth\nonumber\\
&=&-n\Delta\hat\bvarth^T_1(\Sigma^{11})^{-1}\Delta\hat\bvarth_1+o_P(1).
\end{eqnarray}

Now denote $\Phi=(I_q, 0_{q\times s})\Sigma^{-1}(I_q, 0_{q\times s})^T$. It is easy to know that $\Phi=\Sigma^{11}$.
From the proof of Corollary 1, it is known that
$\sqrt n(\hat\bal_1-\bal^{\star}_1)\rightarrow N(0,\sigma^2_1\Phi).$
Thus we obtain that
\begin{eqnarray}\label{eqnA.17}
\mbox{RSS}_0-\mbox{RSS}_1&=&\|\Phi^{-1/2}[\sqrt n(\hat\bal_1-\bal^{\star}_1)]+\sqrt n \Phi^{-1/2} \mb{h_n}\|^2_2+o_P(1).
\end{eqnarray}

On the other hand, we have
\begin{eqnarray*}
\frac{\mbox{RSS}_1}{n-q}&=&\frac{1}{n-q}\|\bY-\bM\hat\bal_{0}-\bX\hat\bal_1\|^2_2=
\frac{1}{n-q}\|\bY-\bM\bal^{\star}_{0}-\bX\bal^{\star}_1\|^2_2\\
&&-2\frac{1}{n-q}(\hat\bvarth-\bvarth_0)^T\dot{L}(\bvarth_0)
+\frac{1}{n-q}(\hat\bvarth-\bvarth_0)^T nD_1(\hat\bvarth-\bvarth_0)\\
&=&I_1-2I_2+I_3.
\end{eqnarray*}
It is easy to know that $I_1\rightarrow \sigma^2_1$, while $I_3\leq
\|\hat\bvarth-\bvarth_0\|^2_2\|D_1\|_2=O_P(s/n)=o_P(1)$. Further note that
\begin{eqnarray*}
I_2\leq \|\hat\bvarth-\bvarth_0\|_2\|
\dot{L}(\bvarth_0)\|_2/(n-q)=O_P((1/n)\sqrt{s/n}\sqrt{ns})=O_P(\frac{s}{n})=o_P(1).
\end{eqnarray*}
In sum, it follows that
\begin{eqnarray}\label{eqnA.18}
\frac{\mbox{RSS}_1}{n-q}=\sigma^2_1+o_P(1).
\end{eqnarray}
As a result, we have
\begin{eqnarray*}
T_n=\frac{\mbox{RSS}_0-\mbox{RSS}_1}{\mbox{RSS}_1/(n-q)}\rightarrow
\chi^2_{q}(n\mb{h_n}^T\Phi^{-1}\mb{h_n}/\sigma^2_1).
\end{eqnarray*}

\subsection{Natural direct and indirect effects}
Under the independence conditions of random errors in the models, the
sequential ignorability assumption \citep{Imai:Keele:Tingley:2010}
holds, and the natural direct and indirect effects can be
identified. As argued by \cite{Imai:Keele:Tingley:2010}, only the
sequential ignorability assumption is needed and neither the
linearity nor the no-interaction assumption is required for the
identification of mediation effects. However, in the situation
with high dimensional mediators, it would be very challenging if
not impossible to make inference about the mediation models
without linearity nor the no-interaction assumption. The linearity
and the no-interaction assumptions are widely adopted in recent
studies about HDMM \citep{Zhang.etal:2016,van Kesteren:Oberski:2019,Zhou:Wang:Zhao:2020}.

To define the natural direct and natural indirect effects, we give some notation
first. Let $y(\bx^*,\mbm^*)$ denote the potential outcome that
would have been observed had $\bx$ and $\mbm$ been set to $\bx^*$ and $\mbm^*$, respectively, and
$\mbm(\bx^*)$ denotes the potential mediator that would have been observed had $\bx$ been
set to $\bx^*$. Following \cite{Imai:Keele:Tingley:2010,Vanderweele:Vansteelandt:2014},
and others, for $\bx=\bx_1$ versus $\bx_0$, the natural direct effect is defined as
$E[y(\bx_1, \mbm(\bx_0))-y(\bx_0,\mbm(\bx_0))].$
While the indirect effect is defined as
$E[y(\bx_1, \mbm(\bx_1))-y(\bx_1,\mbm(\bx_0))].$
Then the total effect $E[y(\bx_1, \mbm(\bx_1))-y(\bx_0,\mbm(\bx_0))]$ is the sum of the natural direct and indirect effect.
\cite{Vanderweele:Vansteelandt:2014} showed that
\begin{align*}
&E[y(\bx_1, \mbm(\bx_0))-y(\bx_0,\mbm(\bx_0))]=\bal_1^T(\bx_1-\bx_0)\\
&E[y(\bx_1, \mbm(\bx_1))-y(\bx_1,\mbm(\bx_0))]=(\Gamma\bal_0)^T(\bx_1-\bx_0).
\end{align*}
Thus $\bal_1$ can be interpreted as the average natural direct effect, and $\bbeta=\Gamma\bal_0$ can
be interpreted as the average natural indirect effect, of a one-unit change in the exposure $\bx$.


\begin{thebibliography}{99}
\expandafter\ifx\csname natexlab\endcsname\relax\def\natexlab#1{#1}\fi

\bibitem[{Abarbanell \& Bushee(1997)}]{Abarbanell:Bushee:1997}
\textsc{Abarbanell, J. S., \& Bushee, B. J.}(1997).
\newblock Fundamental analysis, future earnings, and stock prices.
\newblock \textit{Journal of Accounting Research}, \textbf{35}(1), 1-24.

\bibitem[{Ai et~al.(2021)}]{Ai:2021}
\textsc{Ai, C., Linton, O. \& Zhang, Z.}(2021).
\newblock Estimation and inference for the counterfactual distribution and quantile functions in continuous treatment models.
\newblock \textit{Journal of Econometrics}, https://doi.org/10.1016/j.jeconom.2020.12.009.

\bibitem[{Athey et~al.(2018)}]{Athey:Imbens:Wager:2018}
\textsc{Athey, S., Imbens, G. \& Wager, S.}(2018).
\newblock Approximate residual balancing: Debiased inference of average treatment effects in high dimensions.
\newblock \textit{Journal of the
Royal Statistical Society, Series B} {\bf 80}, 597-623.


\bibitem[{Baron \& Kenny(1986)}]{Baron:Kenny:1986}
\textsc{Baron, R. M. \& Kenny, D. A.}(1986).
\newblock The moderator-mediator variable distinction in social psychological research: Conceptual, strategic, and statistical considerations.
\newblock \textit{Journal of Personality and Social Psychology} \textbf{51}(6), 1173-1182.

\bibitem[{Baker et al.(2020)}]{Baker:2020}
\textsc{Baker, S. R., Bloom, N., Davis, S. J., Kost, K., Sammon, M., \& Viratyosin, T.}(2020).
\newblock The unprecedented stock market reaction to COVID-19.
\newblock \textit{The Review of Asset Pricing Studies}, \textbf{10}(4), 742-758.

\bibitem[{Belloni et~al.(2014)}]{Belloni2014}
\textsc{Belloni, A., Chernozhukov, V. \& Hansen, C.}(2014).
\newblock Inference on treatment effects after selection among high-dimensional controls.
\newblock \textit{The Review of Economic Studies} \textbf{81}(2), 608-650.


\bibitem[{Callen \& Segal(2004)}]{Callen:Segal:2004}
\textsc{Callen, J. L., \& Segal, D.} (2004).
\newblock Do accruals drive firm‐level stock returns? A variance decomposition analysis. \newblock \textit{Journal of Accounting Research}, \textbf{42}(3), 527-560.

\bibitem[{Chakrabortty et~al.(2018)}] {Chakrabortty:Nandy:Li:2018}
\textsc{Chakrabortty, A.}, \textsc{Nandy, P.} \& \textsc{Li, H.}(2018).
\newblock Inference for individual mediation effects and interventional effects in sparse high-dimensional causal graphical models.
\newblock \textit{arXiv}: 1809.10652v1.

\bibitem[{Charnes et al.(1978)}]{Charnes:1978}
\textsc{Charnes, A., Cooper, W. W., \& Rhodes, E.}(1978).
\newblock Measuring the efficiency of decision making units.
\newblock \textit{European Journal of Operational Research}, \textbf{2}(6), 429-444.

\bibitem[{Chen et~al.(2018)}] {Chen.etal:2018}
\textsc{Chen, O. Y., Crainiceanu, C., Ogburn, E. L., Caffo, B. S., Wager, T. D. \& Lindquist, M. A.}
 (2018).
\newblock High-dimensional multivariate mediation with application to neuroimaging data.
\newblock \textit{Biostatistics} \textbf{19}(2), 121-136.

\bibitem[{Chernozhukov et~al.(2021)}] {Chernozhukov:2021}
\textsc{Chernozhukov, V.}, \textsc{Kasahara, H. J.} \& \textsc{Schrimpf, P.}(2021).
\newblock Causal impact of masks, policies, behavior on early covid-19 pandemic in the US.
\newblock \textit{Journal of Econometrics} \textbf{220}(1), 23-62.


\bibitem[{Conti et~al.(2016)}] {Conti:2016}
\textsc{Conti, G., Heckman, J. J. \& Pinto, R.} (2016).
\newblock The effects of two influential early childhood interventions on health and healthy behaviour.
\newblock \textit{The Economic Journal} \textbf{126}(596), 28-65.

\bibitem[{De Vito \& G{\'o}mez(2020)}]{Devito:2020}
\textsc{De Vito, A., \& G{\'o}mez, J. P.}(2020).
\newblock Estimating the COVID-19 cash crunch: Global evidence and policy.
\newblock \textit{Journal of Accounting and Public Policy}, \textbf{39}(2), 106741.

\bibitem[{Derkach et~al.(2019)}] {Derkach:Pfeiffer:Chen:Sampson:2019}
\textsc{Derkach, A., Pfeiffer, R. M., Chen, T. H. \& Sampson, J. N.}(2019).
\newblock High dimensional mediation analysis with latent variables.
\newblock \textit{Biometrics} \textbf{75}(3), 745-756.

\bibitem[{Dimitropoulos \& Asteriou(2009)}]{Dimitropoulos:Asteriou:2009}
\textsc{Dimitropoulos, P. E., \& Asteriou, D.}(2009).
\newblock The value relevance of financial statements and their impact on stock prices. \newblock \textit{Managerial Auditing Journal}, \textbf{24}(3), 248-–265.

\bibitem[{Donald \& Hsu(2014)}] {Donald:2014}
\textsc{Donald, S. G. \& Hsu, Y. C.}(2014).
\newblock Estimation and inference for distribution functions and quantile functions in treatment effect models.
\newblock \textit{Journal of Econometrics} \textbf{178},  383-397.

\bibitem[{Edirisinghe \& Zhang(2007)}]{Edirisinghe:Zhang:2007}
\textsc{Edirisinghe, N. C., \& Zhang, X.}(2007).
\newblock Generalized DEA model of fundamental analysis and its application to portfolio optimization.
\newblock \textit{Journal of banking \& finance}, \textbf{31}(11), 3311-3335.

\bibitem[{Edirisinghe \& Zhang(2008)}]{Edirisinghe:Zhang:2008}
\textsc{Edirisinghe, N. C. P., \& Zhang, X.} (2008).
\newblock Portfolio selection under DEA-based relative financial strength indicators: case of US industries.
\newblock \textit{Journal of the Operational Research Society}, \textbf{59}(6), 842-856.

\bibitem[{Enke \& Thawornwong(2005)}]{Enke:Thawornwong:2005}
\textsc{Enke, D., \& Thawornwong, S.}(2005).
\newblock The use of data mining and neural networks for forecasting stock market returns.
\newblock \textit{Expert Systems with Applications}, \textit{29}(4), 927-940.


\bibitem[{Fama \& French(1993)}]{Fama:French:1993}
\textsc{Fama, E. F., \& French, K. R.} (1993).
\newblock Common risk factors in the returns on stocks and bonds.
\newblock \textit{Journal of Financial Economics}, \textbf{33}, 3–-56.

\bibitem[{Fama \& French(2015)}]{Fama:French:2015}
\textsc{Fama, E. F., \& French, K. R.}(2015).
\newblock A five-factor asset pricing model.
\newblock \textit{Journal of Financial Economics}, \textbf{116}(1), 1-22.

\bibitem[{Fan et~al.(2012)}]{Fan:Guo:Hao:2012}
\textsc{Fan, J., Guo, S. \& Hao, N.}(2012).
\newblock Variance estimation using refitted cross-validation
in ultrahigh dimensional regression.
\newblock \textit{Journal of the
Royal Statistical Society, Series B} \textbf{74}, 37-65.

\bibitem[{Fan \& Li(2001)}]{Fan:Li:2001}
\textsc{Fan, J. \& Li, R.}(2001).
\newblock Variable selection via nonconcave penalized likelihood and its oracle properties.
\newblock \textit{Journal of the American Statistical Association} \textbf{96}, 1348-1360.

\bibitem[{Fan et al.(2020)}]{Fan:Li:Zhang:Zou:2020}
\textsc{Fan, J., Li, R., Zhang, C.-H. \& Zou, H.}(2020).
\textit{Statistical Foundations of Data Science}.
\newblock Chapman and Hall/CRC. Boca Raton, FL.

\bibitem[{Fan \& Lv(2011)}]{Fan:Lv:2011}
\textsc{ Fan, J. \& Lv, J.} (2011).
\newblock Nonconcave penalized likelihood with NP-dimensionality.
\newblock \textit{IEEE Transactions on Information Theory} \textbf{57}, 5467-5484.

\bibitem[{Fan, et~al(2020a)}]{FanY2020a}
\textsc{Fan, Y., Lv, J., Sharifvaghefi, M. and Uematsu, Y.}
(2020a).
\newblock IPAD: stable interpretable forecasting with knockoffs inference.
\newblock \textit{Journal of the American Statistical Association} \textbf{115}, 1822-1834.

\bibitem[{Fan, et~al(2020b)}]{FanY2020b}
\textsc{Fan, Y., Demirkaya, E., Li, G. and Lv, J.} (2020b).
\newblock RANK: large-scale inference with graphical nonlinear knockoffs.
\newblock \textit{Journal of the American Statistical Association} \textbf{115}, 362-379.


\bibitem[{Farrell(1957)}]{Farrell:1957}
\textsc{Farrell, M. J.}(1957).
\newblock The measurement of efficiency productive.
\newblock \textit{Journal of the Royal Statistical Society}, \textbf{120}(3), 253-266.


\bibitem[{Gormsen \& Koijen(2020)}]{Gormsen:2020}
\textsc{Gormsen, N. J., \& Koijen, R. S.}(2020).
\newblock  Coronavirus: Impact on stock prices and growth expectations.
\newblock \textit{The Review of Asset Pricing Studies}, \textbf{10}(4), 574-597.

\bibitem[{Graham et al.(2002)}]{Graham:2002}
\textsc{Graham, Carol M., Mark V. Cannice, \& Todd L. Sayre.}(2002)
\newblock The value‐relevance of financial and non‐financial information for Internet companies.
\newblock \textit{Thunderbird International Business Review}, \textbf{44}(1), 47-70.


\bibitem[{Hassan et al.(2020)}]{Hassan:2020}
\textsc{Hassan, T. A., Hollander, S., van Lent, L., Schwedeler, M., \& Tahoun, A.}(2020).
\newblock Firm-level Exposure to Epidemic Diseases: Covid-19, SARS, and H1N1.
\newblock \textit{National Bureau of Economic Research}, \textbf{No.26971}.


\bibitem[{Hayes(2013)}]{Hayes:2013}
\textsc{Hayes, A. F.} (2013).
\newblock Introduction to Mediation, Moderation, and Conditional Process Analysis: A Regression-based Approach.
\newblock Guilford Press.

\bibitem[{Huang et al.(2019)}]{Huang:2019}
\textsc{Huang, Y., Capretz, L. F., \& Ho, D.} (2019).
\newblock Neural network models for stock selection based on fundamental analysis.
\newblock \textit{In 2019 IEEE Canadian Conference of Electrical and Computer Engineering (CCECE)} (pp. 1-4). IEEE.

\bibitem[{Huang \& Pan(2016)}]{Huang:Pan:2016}
\textsc{Huang, Y. T. \& Pan, W. C.}(2016).
\newblock Hypothesis test of mediation effect in causal mediation model with high-dimensional continuous mediators.
\newblock \textit{Biometrics} \textbf{72}(2), 402-413.


\bibitem[{Huber et al.(2020)}]{Huber:2020}
\textsc{Huber, M., Hsu, Y. C., Lee, Y. Y. \& Lettry, L.}(2020).
\newblock Direct and indirect effects of continuous treatments based
on generalized propensity score weighting.
\newblock \textit{Journal of Applied Econometrics} \textbf{35}(7), 814-840.

\bibitem[{Imai et~al.(2010)}]{Imai:Keele:Tingley:2010}
\textsc{Imai, K., Keele, L.\& Tingley, D.}(2010).
\newblock A general approach to causal mediation analysis.
\newblock \textit{Psychological Methods} \textbf{15}(4), 309-334.

\bibitem[{Imbens(2004)}]{Imbens:2004}
\textsc{Imbens, G.} (2004).
\newblock Nonparametric estimation of average treatment effects under exogeneity: A review.
\newblock \textit{The Review of Economics and Statistics} \textbf{86}(1), 4-29.

\bibitem[{Javanmard \& Montanari(2014)}]{Javanmard2014}
\textsc{Javanmard, A. \& Montanari, A.}(2014).
\newblock Confidence intervals and hypothesis testing for high-dimensional regression.
\newblock \textit{Journal of Machine Learning Research} {\bf 15}(1), 2869-2909.



\bibitem[{Khan \& Khokhar(2015)}]{Khan:Khokhar:2015}
\textsc{Khan, M. N., \& Khokhar, I.} (2015).
\newblock Quarterly Journal of Econometrics Research.
\newblock \textit{Quarterly Journal of Econometrics Research}, \textbf{1}(1), 1-12.

\bibitem[{Lee et al.(2019)}]{Lee:2019}
\textsc{Lee, T. K., Cho, J. H., Kwon, D. S., \& Sohn, S. Y.}(2019).
\newblock Global stock market investment strategies based on financial network indicators using machine learning techniques.
\newblock \textit{Expert Systems with Applications}, \textbf{117}, 228-242.



\bibitem[{Mackinnon(2008)}]{Mackinnon:2008}
\textsc{Mackinnon, D. P.} (2008).
\newblock Introduction to Statistical Mediation Analysis.
\newblock Routledge

\bibitem[{Preacher(2015)}]{Preacher:2015}
\textsc{Preacher, K. J.} (2015).
\newblock Advances in mediation analysis: A survey and synthesis of new developments.
\newblock \textit{Annual Review of Psychology} \textbf{66}(1),  825-852.

\bibitem[{Preacher \& Hayes(2008)}]{Preacher:Hayes:2008}
\textsc{Preacher, K. J. \& Hayes, A. F.} (2008).
\newblock Asymptotic and resampling strategies for assessing and comparing indirect effects in multiple mediator models.
\newblock \textit{Behavior Research Methods} \textbf{40}(3), 879-891.


\bibitem[{Ramelli \& Wagner(2020)}]{Ramelli:2020}
\textsc{Ramelli, S., \& Wagner, A. F.} (2020).
\newblock Feverish stock price reactions to COVID-19.
\newblock \textit{The Review of Corporate Finance Studies}, \textbf{9}(3), 622-655.


\bibitem[{Shi et~al.(2019)Song, Chen \&
  Li}]{Shi:Song:Chen:Li:2019}
\textsc{ Shi, C., Song, R., Chen, Z., \& Li, R. }
 (2019).
\newblock Linear hypothesis testing for high-dimensional generalized linear model.
\newblock \textit{Annals of Statistics} \textbf{47}(5), 2671-2703.

\bibitem[{Song et~al.(2020)}]{Song.etal:2020}
\textsc{Song, Y., Zhou, X., Zhang, M., Zhao, W.,
Liu, Y., Kardia, S. L. R., Roux, A. V. D., Needham, B. L., Smith, J. A. \& Mukherjee, B.}(2020).
\newblock Bayesian shrinkage estimation of high dimensional
causal mediation effects in omics studies.
\newblock \textit{Biometrics} \textbf{76}(3), 700-710.

\bibitem[{Sun \& Zhang(2013)}]{Sun:Zhang:2013}
\textsc{Sun, T. \& Zhang, C. H.} (2013).
\newblock Scaled sparse linear regression.
\newblock \textit{Biometrika} \textbf{99}(4), 879-898.

\bibitem[{Ten Have \& Joffe(2010)}]{Ten:Joffe:2010}
\textsc{Ten Have, T. \& Joffe, M.} (2010).
\newblock A review of causal estimation of effects in mediation analyses.
\newblock \textit{Statistical Methods in Medical Research} \textbf{21}(1), 77-107.

\bibitem[{Thorbecke (2020)}]{Thorbecke:2020}
\textsc{Thorbecke, W.}(2020)
\newblock The Impact of the COVID-19 Pandemic on the US Economy: Evidence from the Stock Market.
\newblock \textit{Journal of Risk and Financial Management}, \textbf{13}(10), 233.

\bibitem[{van de Geer et~al.(2014)}]{van de Geer:Buhlmann:Ritov:Dezeure:2014}
\textsc{van de Geer, S., Buhlmann, P., Ritov, Y. \& Dezeure, R.}
 (2014).
\newblock On asymptotically optimal confidence regions and tests for high-dimensional models.
\newblock \textit{Annals of Statistics} \textbf{42}(3), 1166-1202.

\bibitem[{van Kesteren \& Oberski(2019)}]{van Kesteren:Oberski:2019}
\textsc{van Kesteren, E. J. \& Oberski, D. L.}(2019).
\newblock Exploratory mediation analysis with many potential mediators.
\newblock \textit{Structural Equation Modeling: A Multidisciplinary Journal} \textbf{26}(5), 710-723.

\bibitem[{Vanderweele(2015)}]{Vanderweele:2015}
\textsc{Vanderweele, T. J.}(2015).
\newblock Explanation in Causal Inference: Methods for Mediation and Interaction.
\newblock  Oxford University Press. New York.

\bibitem[{Vanderweele \& Vansteelandt(2014)}] {Vanderweele:Vansteelandt:2014}
\textsc{Vanderweele, T. J. \& Vansteelandt, S.}(2014).
\newblock Mediation analysis with multiple mediators.
\newblock \textit{Epidemiologic Methods} \textbf{2}(1), 95-115.

\bibitem[{Wang et~al.(2013)Kim \&
  Li}]{Wang:Kim:Li:2013}
\textsc{Wang, L., Kim, Y. \& Li, R.}(2013).
\newblock Calibrating non-convex penalized regression in ultra-high dimension.
\newblock \textit{Annals of Statistics} \textbf{41}(5), 2505-2536.

\bibitem[{Wang et~al.(2012)Wu \&
  Li}]{Wang:Wu:Li:2012}
\textsc{ Wang, L., Wu, Y. \& Li, R.} (2012).
\newblock Quantile regression for analyzing heterogeneity in ultra-high dimension.
\newblock \textit{Journal of the American Statistical Association} \textbf{107}(497), 214-222.

\bibitem[{Zhang et al.(2020)}]{Zhang:2020}
\textsc{Zhang, D., Hu, M., \& Ji, Q.}(2020).
\newblock Financial markets under the global pandemic of COVID-19. \newblock \textit{Finance Research Letters}, \textbf{36}, 101528.

\bibitem[{Zhang et~al.(2016)Zheng, Holmes \&
  Stephens}]{Zhang.etal:2016}
\textsc{Zhang, H., Zheng, Y., Zhang, Z., Gao, T., Joyce, B., Yoon, G., Zhang, W., Schwartz, J., Just, A., Colicino, E., Vokonas, P., Zhao, L., Lv, J., Baccarelli, A., Hou, L. \& Liu, L.}(2016).
\newblock Estimating and testing high-dimensional mediation effects in epigenetic studies.
\newblock \textit{Bioinformatics} \textbf{32}(20), 3150-3154.

\bibitem[{Zhang \& Zhang(2014)}]{Zhang:Zhang:2014}
\textsc{Zhang, C. H. \& Zhang, S. S.}(2014).
\newblock Confidence intervals for low dimensional parameters in high dimensional linear models.
\newblock \textit{Journal of the
Royal Statistical Society, Series B} \textbf{76}, 217-242.


\bibitem[{Zhao et~al.(2020)Linqduist \&
  Caffo}]{Zhao:Linqduist:Caffo:2020}
\textsc{Zhao, Y. Linqduist, M. A., \& Caffo, B.S.}
 (2020).
\newblock Sparse principal component based high-dimensional mediation analysis.
\newblock \textit{Computational Statistics and Data Analysis} \textbf{142}, 106835.

\bibitem[{Zhou et~al.(2020)Wang \& Zhao}]{Zhou:Wang:Zhao:2020}
\textsc{Zhou, R. X., Wang, L. W., \& Zhao, S. H.}(2020).
\newblock Estimation and inference for the indirect effect in high-dimensional linear mediation models.
\newblock \textit{Biometrika} \textbf{107}(3), 573-589.

\bibitem[{Zou \& Li(2008)}]{Zou:Li:2008}
\textsc{ Zou, H. and Li, R.} (2008).
\newblock One-step sparse estimates in nonconcave penalized likelihood models.
\newblock \textit{Annals of Statistics} \textbf{36}(4), 1509-1533.



\end{thebibliography}
\end{document}